\documentclass[twocolumn]{aastex631}

\usepackage{multirow}
\usepackage[]{hyperref}
\usepackage{subfigure}
\usepackage{amsmath}
\usepackage{enumitem}
\usepackage{todonotes}

\hypersetup{
    colorlinks=true,
    linkcolor=blue,
    filecolor=magenta,      
    urlcolor=cyan,
    pdftitle={Improving Pulsar Timing Precision with Single Pulse Fluence Clustering},
    pdfpagemode=FullScreen,
    }

\begin{document}

\title{Improving Pulsar Timing Precision with Single Pulse Fluence Clustering}

\author[0000-0002-5176-2924]{Sofia V. Sosa Fiscella}
\affiliation{School of Physics and Astronomy, Rochester Institute of Technology, Rochester, NY 14623, USA}
\affiliation{Laboratory for Multiwavelength Astrophysics, Rochester Institute of Technology, Rochester, NY 14623, USA}
\author[0000-0003-0721-651X]{Michael T. Lam}
\affiliation{SETI Institute, 339 N Bernardo Ave Suite 200, Mountain View, CA 94043, USA}
\affiliation{School of Physics and Astronomy, Rochester Institute of Technology, Rochester, NY 14623, USA}
\affiliation{Laboratory for Multiwavelength Astrophysics, Rochester Institute of Technology, Rochester, NY 14623, USA}
\author[0000-0001-7697-7422]{Maura A. McLaughlin}
\affiliation{Department of Physics and Astronomy, West Virginia University, P.O. Box 6315, Morgantown, WV 26506, USA}
\affiliation{Center for Gravitational Waves and Cosmology, West Virginia University, Chestnut Ridge Research Building, Morgantown, WV 26505, USA}




\begin{abstract}
Traditional pulsar timing techniques involve averaging large numbers of single pulses to obtain a high signal-to-noise (S/N) profile, which is matched to a template to measure a time of arrival (TOA). However, the morphology of individual single pulses varies greatly due to pulse jitter. Pulses of different fluence contribute differently to the S/N of the pulse average. Our study proposes a method that accounts for these variations by identifying a range of ``states" and timing each separately. We selected two 1-hour observations of PSR J2145$-$0750, each in a different frequency band with the Green Bank Telescope. We normalized the pulse amplitudes to account for scintillation effects and probed different excision algorithms to reduce radio-frequency interference. We then measured four pulse parameters (amplitude, position, width, and energy) to classify the single pulses using automated clustering algorithms. For each cluster, we calculated an average pulse profile and template and used both to obtain a TOA and TOA error. Finally, we computed the weighted average TOA and TOA error, the latter as a metric of the total timing precision for the epoch. The TOA is shifted relative to the one obtained without clustering, and we can estimate the shift with this weighting using the same data. For the $820$-MHz and 1400-MHz bands, we obtained TOA uncertainties of $0.057~\mu$s and $0.46~ \mu$s, compared to $0.066~\mu$s and $0.74~\mu$s when no clustering is applied. We conclude that tailoring this method could help improve the timing precision for certain bright pulsars in NANOGrav's dataset.

\end{abstract}



\section{Introduction} \label{sec:intro}

\subsection{TOA error due to single pulse variability}

Pulsar timing is a technique that involves using observations of radio pulsar pulse profiles to calculate pulse arrival times, which are then compared with mathematical models that incorporate a wide variety of astrophysical phenomena. The many merits of this technique include characterizing the orbits of binary systems \citep{2016ApJ...832..167F}, enabling tests of general relativity \citep{2006Sci...314...97K}, constraining nuclear equations of state \citep{2021ApJ...915L..12F, 2020NatAs...4...72C, 2013ApJ...762...94D, 2013Sci...340..448A}, and detecting planetary-mass companions \citep{1992Natur.355..145W}. Moreover, by monitoring variations in pulse times of arrival (TOAs) from an array of the most stable millisecond pulsars (MSPs), evidence has been detected for a signal consistent with nanohertz frequency gravitational radiation from a population of supermassive black hole binaries \citep{2023ApJ...951L...8A, 1983ApJ...265L..39H, 1979ApJ...234.1100D}.

As pulsars are generally weak radio sources, pulsar timing applications require the addition of several single pulses. This technique, known as folding, utilizes pulsars' remarkably stable rotation to time-average hundreds of thousands of single pulses. As a result, the background noise is decreased while the single pulses are added in phase, increasing the signal-to-noise (S/N) ratio. The average profile formed by averaging a large number of single pulses converges to a shape that appears to be epoch-independent \citep[see][]{craft_1970, 1973ApJ...182..245B, 1992ApJ...385..273P, 2012A&A...543A..66H, 2016yCat..35860092P}.

However, nearly every pulsar observed with high sensitivity shows intrinsic single-pulse stochastic variability in excess of that expected from radiometer noise \citep[see][]{1985ApJS...59..343C, 1990ApJ...349..245C, 10.1111/j.1365-2966.2011.20041.x, 10.1093/mnras/stu1213, Dolch_2014, 2012ApJ...761...64S}. This includes variations in amplitude and phase that are correlated from pulse to pulse (such as the drifting sub-pulse phenomenon) and variations that are uncorrelated from pulse to pulse.  When averaged to form a pulse profile, this variability causes the underlying pulse shape to differ from that of the template. This difference biases the measurements of arrival times, contributing to what is called ``jitter noise'' in the TOAs \citep{Shannon_2010}. The jitter TOA error is independent of S/N, so it cannot be mitigated with improved observing backends.


Given the importance of precise timing for PTA experiments, several studies have attempted to quantify and mitigate the presence of pulse jitter in MSPs. \cite{2012ApJ...761...64S} showed that intrinsic single-pulse variations in amplitude, shape, and pulse phase for the MSP J1713$+$0747 are largely responsible for the excess in timing errors for that pulsar. Moreover, they found that brighter single pulses tend to have earlier arrival times. Most importantly, they investigated two methods for correcting TOAs due to single-pulse variations: multi-component template fitting and principal component analysis. However, both algorithms were unsuccessful at improving the precision of arrival times using pulse-shape information.

On the other hand, a growing body of evidence suggests that MSPs undergo short-scale mode changes:
\begin{itemize}
    \item \cite{2014MNRAS.441.3148O} found that for PSR J0437$-$4715 the average profile changes as a function of single-pulse S/N with the highest S/N pulses giving the narrowest pulse average. They also found that removing the lowest S/N pulses from the TOA analysis reduced the overall RMS timing residual.
    \item  \cite{2014MNRAS.443.1463S} observed that many pulsars show the same changes in average pulse profile as a function of the S/N of its single pulses. Furthermore, they constructed integrated pulse profiles using only the 100 most energetic single pulses for each of their pulsars, since the bright, narrow pulses typically emanated from a small region in the pulse phase. For PSR J0437$-$4715, they found that the width of the average of the brightest pulses is 80 $\mu$s, which is significantly smaller than the 140 $\mu$s width of all of the pulses. However, they did not pursue any timing analysis with the average of the brightest pulses.
    \item Finally, \cite{2015MNRAS.452..607K} treated the pulse-to-pulse variability as arising from a finite family of pulse shapes. They classified single pulses into different categories based on their similarity (or dissimilarity) to the pulse template. Based on this classification, they devised timing algorithms that optimize the pulse template by creating a template basis that describes the single-pulse variations. In doing so, they reduced jitter error in an observation of J0835$-$4510 (Vela) by 30–40\%. Interestingly, they also find a strong correlation between the peak intensity of a pulse and the phase at which that peak falls.
\end{itemize}

\noindent This body of work emphasizes the need for an in-depth study into single-pulse variability and how to account for it in the pulsar timing process.

\subsection{TOA Errors from Additive Noise}

In the present work, we assess the merit of a method for improving TOA precision based on single-pulse analysis that, unlike \cite{2012ApJ...761...64S}'s jitter-based approach, attempts to mitigate the error from additive noise. This error arises from single pulses having radiometer noise with a Gaussian probability density function that is additive when averaging them to create the integrated pulse profile. In turn, template matching yields TOA uncertainties that depend on the S/N of the average pulse \citep{2010arXiv1010.3785C}.
 
Traditional pulsar timing techniques assume that the pulse intensity at a given observing frequency $\nu$ can be modeled as a function of time as:
\begin{equation}\label{eq:temp}
    I(t|\nu) = S(t, \nu) \sigma_{\mathrm{n}}(\nu) U(t-t_{0}|\nu) + n(t|\nu)
\end{equation}
\noindent where $U(t)$ is the pulse template shape, $n(t)$ is additive noise with RMS amplitude given by $\sigma_{n}$, $t_{0}$ is the TOA, and $S$ is the S/N of the pulse profile (peak to RMS off-pulse). In that case, the signal model for an integrated pulse profile is given by:

\begin{equation}\label{eq:sigma_TOA}
    \sigma_\mathrm{S/N} (t | \nu) = \frac{W_\mathrm{eff}(\nu)}{S(t, \nu) \sqrt{N_\phi}}
\end{equation}

\noindent where $W_\mathrm{eff}$ is the effective pulse width (i.e., the width of a top-hat pulse with the same area), and $N_\phi$ is the number of phase bins across the pulse \citep[e.g.,][]{Dolch_2014}. The decrease in TOA uncertainty with $S$ is the main reason why integrated pulse profiles are used for timing purposes. The number of single pulses that are needed to be able to achieve a stable profile that can be matched to a template is given by the profile stabilization timescale, and it is different for each pulsar \citep[e.g.:][]{2011AAS...21713907K, TeixeiraProfileSI}.

In this work, we follow \cite{2014MNRAS.441.3148O}'s scheme of filtering single pulses based on their S/N. The authors sought to reduce the RMS timing residual by removing the lowest-S/N single pulses from the integrated pulse profile average. We found that such a scheme can potentially yield reduced TOA uncertainties for short observations ($\sim 1000$ single pulses). However, when a larger number of single pulses is available, we found that the improvement in S/N from single-pulse averaging outweighs the reduction in $W_\mathrm{eff}$ when removing the lowest-S/N pulses. 

Instead of using only the brightest single pulses in an observation, we resume \cite{2015MNRAS.452..607K}'s idea that variations in the single pulse shape correspond to shifts between a range of pulse states, each with a characteristic fluence and integrated pulse profile. For the single pulses corresponding to the same state, the observed shape variability, while stochastic, is not without memory but rather correlated in time. From this perspective, states with different fluences weigh differently when contributing to the S/N of the epoch's pulse average. Therefore, assigning weights based on the fluence portrait when computing the epoch-TOA could potentially yield improved timing precision.

In \cite{2014MNRAS.443.1463S}, the authors found that the single pulses from MSPs with the highest intrinsic energy average to a different shape than when averaging all the pulses. Similarly, we will assume that single pulses belonging to different states average to different pulse shapes that can be modeled as:

\begin{equation}
    I(t|\nu, \Vec{s}) = S(\Vec{s}) \sigma_{\mathrm{n}}(\nu) U(t-t_{0}|\nu, \Vec{s}) + n(t |\nu)
\end{equation}

\noindent where $\Vec{s}$ is a set of parameters (such as pulse energy, S/N, $W_\mathrm{eff}$, etc) that characterize each pulse state. Eq.~\ref{eq:sigma_TOA} can then be rewritten as:

\begin{equation}\label{eq:new_sigma_TOA}
    \sigma_\mathrm{TOA} (S | \nu, \Vec{s}) = \frac{W_\mathrm{eff}(\nu, \Vec{s})}{S(\Vec{s}) \sqrt{N_\phi}}
\end{equation}

From this perspective, conventional timing techniques involve averaging along the $\Vec{s}$-axis to increase $S$. However, we propose that for some pulse states $\Vec{s}$ the decrease in $W_\mathrm{eff}$ that was observed by \cite{2014MNRAS.443.1463S} can overcome the increase in $S$ when averaging a larger number of single pulses and, therefore, provide a smaller $\sigma_\mathrm{TOA}$, as proved by \cite{2014MNRAS.441.3148O} for a smaller number of single pulses.

For the sake of simplicity, we will assume discrete and disjoint pulse states $\Vec{s}_1$, $\Vec{s}_2$, $\Vec{s}_3$, $\ldots$. In the appendix, we consider a theoretical treatment of the timing implications of separating pulses belonging to one of two states: a high or low fluence state. However, in our main work, we do not restrict our analysis to a specific number of fluence-only states. Instead, we assume a range of pulse states, and that all single pulses belonging to the same state share similar morphological features such as energy, width, amplitude, S/N, etc. Under these assumptions, state assignment can be performed utilizing automated clustering techniques based on these single pulse features. These techniques are more sophisticated than manually binning pulses in the various dimensions and can potentially identify groupings in the pulse parameter phase space. We will show later in our work that the choice of algorithm does not change the overall impact of pulse clustering. Once the pulse states have been identified by a given algorithm, we calculate the average pulsar profile of each state which can then be compared to a pulse template to produce a TOA and its corresponding $\sigma_\mathrm{S/N}$.

In Sec.~\ref{sec:observations} we provide information on the data collection and reduction methods for the observations we analyzed in this work, as well as the corresponding corrections for scintillation. In Sec.~\ref{sec:data_analysis} we describe the quantification of the pulse features and the clustering algorithms we used to classify the single pulses. In Sec.~\ref{sec:results}, we summarize the main results, including the TOA error attained with each clustering method. In Sec.~\ref{sec:conclusions} we discuss the implications of this work. Processed data products presented here are publicly available\footnote{See \url{https://github.com/sophiasosafiscella/highfluencetiming} for a living version of the code, and \url{https://zenodo.org/records/10463269} for a frozen version.} as of the date this work is published.


\section{Observations and Data Processing}\label{sec:observations}

We will now describe the observations used in this work, how they were reduced, and how we accounted for the modulations in pulse intensity introduced by interstellar scattering/scintillation.

\subsection{Dataset}

In this work, we analyzed observations of PSR J2145$-$0750. Discovered by \cite{1994ApJ...425L..41B}, this low-mass binary pulsar is in a nearly circular orbit, with an orbital period of 6.8 days \citep{2004A&A...426..631L}. It has a spin period of $16.05$ ms, which is considerably longer than the average period of $\sim 4$ ms of this type of binary pulsar, and a distinctive pulse profile with two components separated by 0.20 of the pulse period in phase \citep{1994ApJ...425L..41B}. Most importantly, this bright pulsar exhibits high flux densities, of $14.25$ mJy at $800$ MHz and $4.85$ mJy at $1400$ MHz \citep{2021ApJS..252....4A}. As a result, it is one of the few MSPs whose single pulses have been characterized. Its individual pulses were first detected by \cite{2003A&A...407..273E}; however, only $\sim 100$ pulses were detected and the statistics of the distribution of pulse energies were not explored.

\begin{deluxetable}{lcc}
\tabletypesize{\scriptsize}
\tablewidth{0pt} 
\tablecaption{\centering Observing frequency bands.}\label{tab:observing_frequencies}
\tablehead{& $820~\mathrm{MHz}$ & $1400~\mathrm{MHz}$ \\} 
\startdata 
Bandwidth (MHz) & $200$ & $800$ \\
Central frequency (MHz) & $820.78125$ & $1499.21875$  \\
Observing time (s) & 3479.4160128 & 3586.08273408 \\
Number of single pulses & 216 704 & 219 392 \\
Mean single pulse S/N$^a$ & 4.971 & 3.890 \\
\enddata
\tablecomments{\\ $^a$S/N defined as the ratio of the pulse peak amplitude to the off-pulse noise mean.}
\end{deluxetable}

Observations of PSR J2145$-$0750 were recorded in an approximately 2 hours-long session on February 1st, 2017, using the 100-m Green Bank Telescope (GBT) radio-telescope of the Green Bank Observatory in West Virginia, USA. The radio receivers used for this observation cover two frequency bands: the $820$-MHz band and the $1400$-MHz (L) band. The source was observed for approximately 1 hour in each frequency band. While the 820-band bandwidth is smaller (200 versus 800 MHz), the pulsar is observed to be brighter and the single-pulse S/N is higher at those frequencies (see Table~\ref{tab:observing_frequencies}). Moreover, the frequencies corresponding to the L-band are more heavily affected by radio-frequency interference (RFI).

The data were collected in ``search'' mode where intensity and polarization sampling occurred at a rate of 10.24~$\mu$s in each frequency channel with coherent dedispersion applied. We used the \texttt{DSPSR} package \citep{2011PASA...28....1V} to split the time series into individual time sub-integrations of one pulse each with 512 phase bins (31.3~$\mu$s resolution) using the NANOGrav 12.5-yr timing model file to phase align the profiles \citep{2020ApJ...905L..34A}. We then calibrated the profiles in polarization using \texttt{PSRCHIVE} \citep{2004PASA...21..302H}. Before each of the observations, a noise diode signal was observed so that differential gain and phase offset corrections were applied before we summed across the polarization channels. No absolute flux calibrations were performed. The end product of the observation was a set of total-intensity pulse profiles for a series of $n_{\mathrm{chan}} = 128$ (820-band) or $512$ (L-band) frequency channels, resulting in a channel bandwidth of $1.5625$~MHz. In the pre-processing stage, all the intensity single pulse profiles were averaged across the polarization channels, the average intensity of the off-pulse baseline was subtracted, and the single pulses were phase-shifted so that the main pulse window (see Sec.~\ref{subsec:features}) is centered in the middle of the data array.

To account for changes in pulse shape and intensity due to the interstellar medium, we will assume that all chromatic delays have been perfectly removed or are negligible over each narrowband channel. These include the dispersive delay from DM, scattering, and frequency-dependent pulse profile evolution. We also assume that the signal polarization has been calibrated perfectly. Further details about the observations, their calibration, and data reduction can be found in NANOGrav's 12.5-year \cite{2020ApJ...905L..34A} and its earlier dataset papers \citep{2016ApJ...821...13A, Lam_2016}. Under these assumptions, we model pulse shapes $I(t,\nu,\phi)$ as a function of phase $\phi$, centered on time $t$ and in a sub-band centered on frequency $\nu$.


\subsection{Corrections for Scintillation}

In addition to jitter and additive noise, a third source of TOA variance in short scales is changes in the interstellar impulse response from multipath scattering, which depends strongly on radio frequency \citep{1990ApJ...349..245C}. This results in a pulse broadening function (PBF) caused by diffractive interstellar scattering/scintillation (DISS). DISS will generally result in intensity modulation on timescales of minutes to hours, depending on observing frequency, DM, and direction. The shape perturbation is correlated over a time equal to the diffractive timescale $t_{d}$ and a frequency range equal to the scintillation bandwidth, $\nu_{d}$ \citep{Hemberger_2008}. The DISS timescale and bandwidth vary strongly with observing frequency, approximately as $\nu^{-6/5}$ and $\nu^{-22/5}$, respectively. 

\begin{figure}[h]
\centering
\subfigure[820-band]{\includegraphics[width=8.75cm]{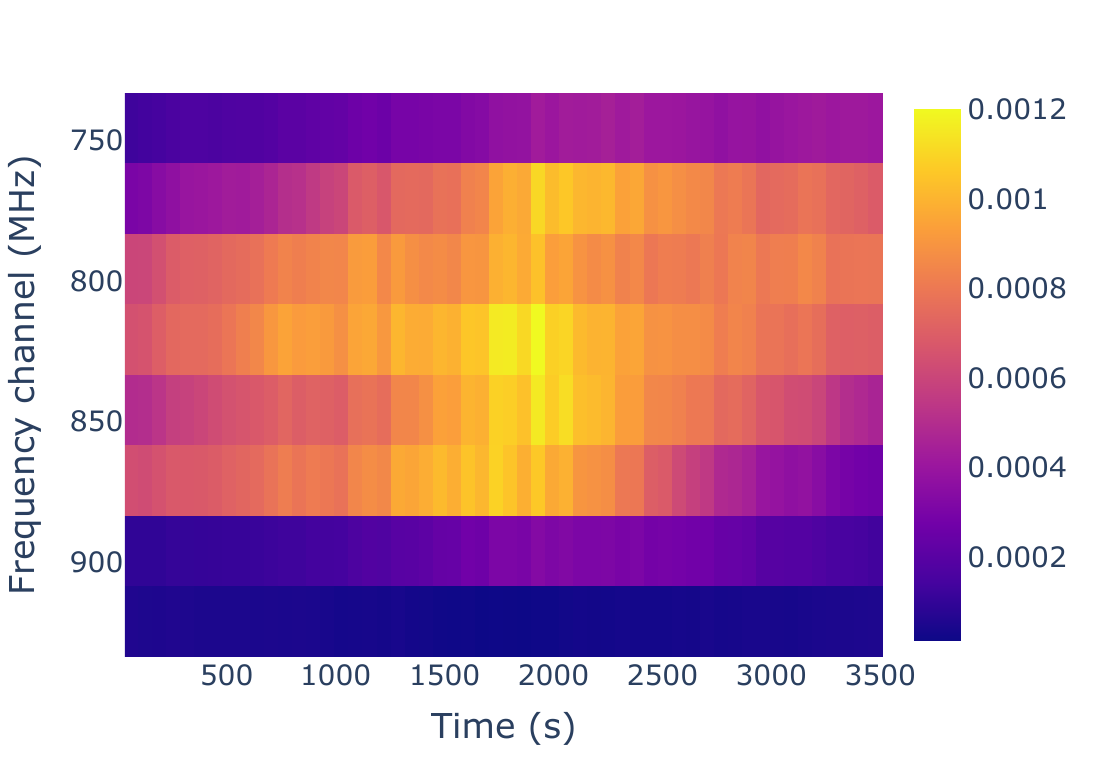}\label{subfig:820-band}}
\centering
\subfigure[L-band]{\includegraphics[width=8.75cm]{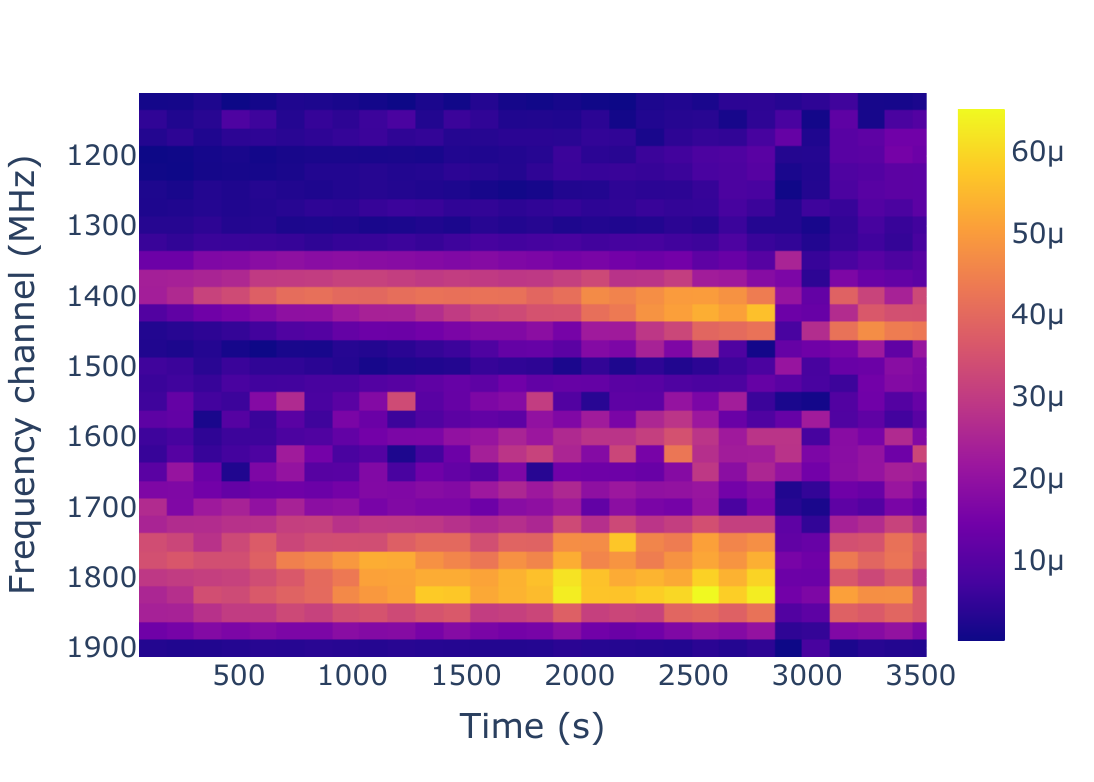}\label{subfig:L-band}}
   
\caption{Dynamic spectrum of the 820-band and the L-band. These were constructed by averaging the observation to $54$ ($\sim$1.1 minutes) or $30$ ($\sim$2 minutes) time sub-integrations, respectively, with $8$ or $32$ frequency channels, and $N_{\phi}=512$ phase bins in both cases. For each $(t, \nu)$ pair, we used the scale factor $b = S \sigma_{0}$ (see Eq.~\ref{eq:temp}), which results from matching the pulse profile to a template, as a measure of the relative pulse amplitude. Since the pulse profiles are not flux-calibrated, the scale factor has arbitrary units.
\label{fig:dynamic_spectrum}}
\end{figure}

As a result of DISS, the time-frequency plane (known as the \textit{dynamic spectrum}) will be made up of independent intensity fluctuations called \textit{scintles} in a time-frequency resolution cell, with a characteristic time $t_{d}$ and frequency scales $\nu_{d}$. The scintillation structure is related to the temporal broadening of pulses, resulting in a time delay \citep{2010arXiv1010.3785C}. 

Because of the intensity fluctuations introduced by the scintles, some low-fluence single pulses will see their amplitudes artificially amplified. Therefore, in order to isolate the single pulses with intrinsic high amplitude, this effect must be mitigated before any single pulse classification can take place. In this work, we accounted for DISS effects with a three-step process:

\begin{enumerate}
    \item First, we used \texttt{PyPulse} \citep{2017ascl.soft06011L} to average the data along the frequency and time, in short integrations longer than the pulse period. We obtained $54$ (820-band) or $30$ (L-band) time sub-integrations with $n_{\mathrm{chan}}=8$ (820-band) or $32$ (L-band) frequency channels and $N_{\phi}=512$ phase bins per single pulse.
    \item We used \texttt{PyPulse}'s \texttt{fitPulses} function to cross-correlate the pulse profiles generated in the previous step with a template profile \citep[e.g.:][]{Lommen_2013}. The template creation procedure is described in \cite{Demorest_2013}; in short, for each epoch and each receiver, the observations of  J2145$-$0750 in the NANOGrav 12.5-year dataset were signal-to-noise-weighted summed and then de-noised via wavelet decomposition and thresholding of the wavelet coefficients (as implemented in the \texttt{PSRCHIVE} function \texttt{psrsmooth}). The scale factor $b = S \sigma_{0}$ (see Eq.~\ref{eq:temp}) that results from correlating each pulse profile with the template was used as a measure of the pulse relative amplitude. We then used those values to construct a dynamic spectrum for each frequency band, presented in Fig.~\ref{fig:dynamic_spectrum}.
    \item Finally, for each single pulse in the original observation, we divided the intensity $I(t,\nu,\phi)$ by the value of the dynamic spectrum in the corresponding $(t,\nu)$. In doing so, we obtained a new dataset normalized by the dynamic spectrum and corrected by the effects of interstellar scintillation.
\end{enumerate}

Next, we averaged the observation in frequency. This process usually involves performing a weighted average where each frequency channel is weighted by its S/N for a given time sub-integration. However, in normalizing by the dynamic spectrum, we artificially modified the RMS noise and, therefore, the corresponding weights. Consequently, we computed new frequency-channel weights as $w_{i}=1/\sigma_{i}^2$ where $\sigma_{i}$ is the RMS of the off-pulse noise. We then used these weights to average in frequency and obtain $I(t,\phi)$.

\section{Single Pulse Clustering}\label{sec:data_analysis}


In this section, we first describe the single-pulse clustering schemes we will use in our analysis. Then we outline an algorithm for calculating TOA measurements and uncertainties for an ensemble of clusters and then weight-averaging them. Finally, we discuss the impact of different RFI mitigation routines.

\subsection{Pulse features}\label{subsec:features}

\begin{figure}[ht]
\epsscale{1.3}  
\plotone{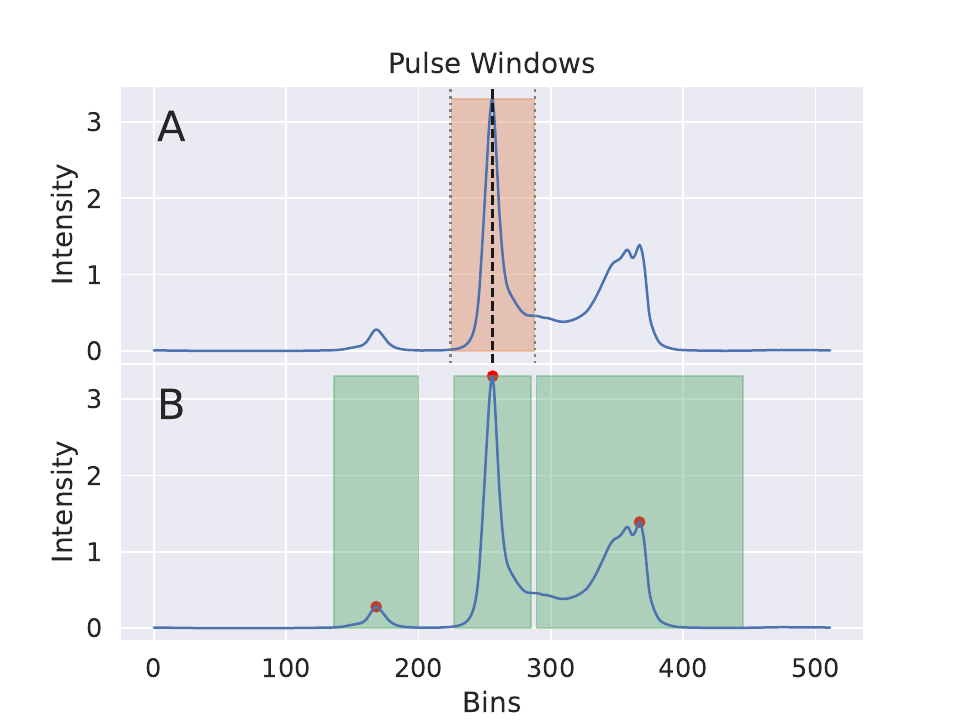}
\caption{Overlapped on the pulsar template, in panel A we present the main pulse window (shaded in orange) that was used for finding the pulse amplitude, width, and position. In panel B we present the pulse component peaks (in red) and the component windows (shaded in green) that were used for finding the pulse energy.
\label{fig:windows}}
\end{figure}

\begin{figure}[ht]
\epsscale{1.2}  
\plotone{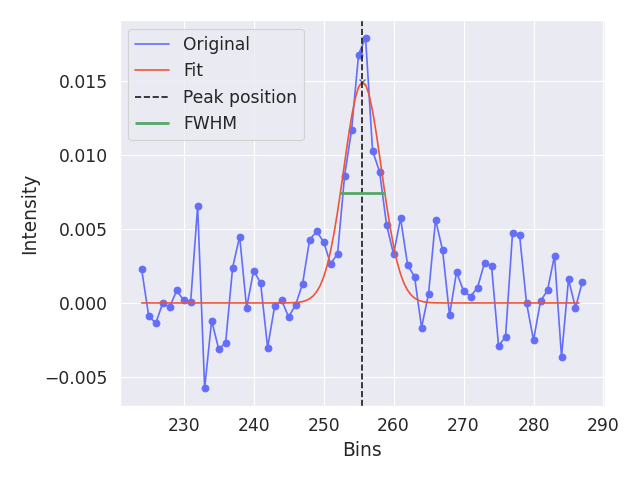}
\caption{Example of a single pulse information inside a main pulse window (see Fig.~\ref{fig:windows}). In blue we present the single pulse and in orange the Gaussian curve that was fit to this pulse to obtain the pulse position (dashed vertical line in black) and full width at half maximum (horizontal green line).
\label{fig:pulse_fit}}
\end{figure}

We classified the single pulses into fluence states based on their morphology. Given the extensive single-pulse variability, in a few single pulses dominated by background noise the intensity maximum in the pulse window would not coincide with the main pulse component. To account for such cases, we constructed a \textit{main component window}. This was created by fitting the position of the template's peak and setting a window around it with a width equal to $12.5\%$ of the template's phase bins. As a result, we obtained a window covering phase bins $[224, 288]$. Similarly, we identified the three pulse components using \texttt{Astropy} \citep{astropy:2013, astropy:2018, astropy:2022} and created component windows around their peaks with a width equal to 2.5 times the component width (see Fig.~\ref{fig:windows}). Since the pulse peak looks jagged, we smooth it by performing a least-squares Gaussian fit to the points inside the main component window (see panel A in Fig.~\ref{fig:pulse_fit}). We then computed four pulse features:

\begin{itemize}
    \item A pulse \textbf{amplitude}, given by the maximum intensity of the main pulse component after subtracting the pulse baseline, which is calculated as the mean of the off-pulse window. Since our observations are not flux-calibrated, the resulting amplitudes will have arbitrary units. However, a flux correspondence could be established by considering that the integrated pulse profiles peak amplitudes, in arbitrary units, of 3.37 ($820$-band, see Fig.~\ref{fig:windows}) and 3.65 (L-band) must correspond to the calibrated flux densities of $14.25$ mJy at $800$ MHz and $4.85$ mJy at $1400$ MHz \citep{2021ApJS..252....4A}.
    \item A pulse \textbf{width}, given by the full width at half maximum of the fitted Gaussian function, in units of phase bins.
    \item A pulse \textbf{position}, given by the position of the center of the peak of the fitted Gaussian function, in units of phase bins.   
    \item A pulse \textbf{energy}, measured as the area under the curve inside the component windows (with the pulse baseline already subtracted). In the case of PSR~J2145$-$0750 using 512 phase bins, these windows cover the ranges $[135, 201]$, $[226, 286]$, and $[288, 446]$ (see panel B in Fig.~\ref{fig:pulse_fit}). As such, the energy will have units of phase bins $\times$ amplitude.
\end{itemize}


\begin{figure}[ht]
\hspace*{-0.7cm}
\centering
\epsscale{1.5}
\plotone{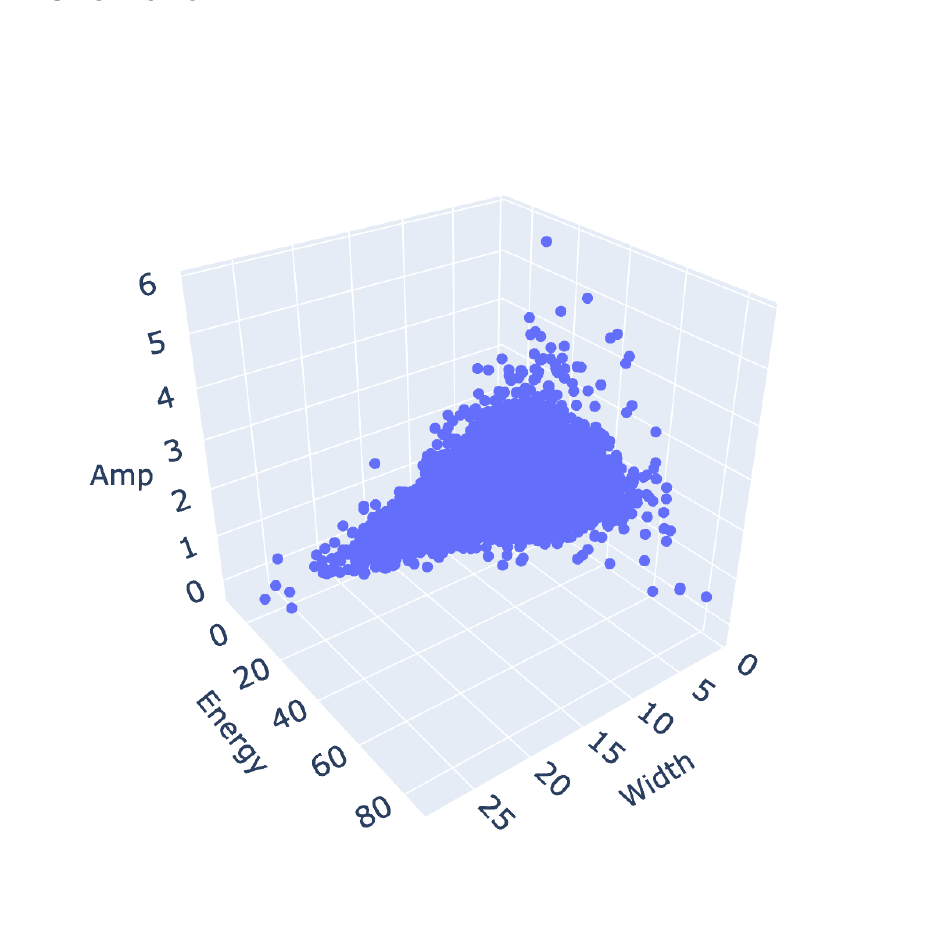}
\caption{3-dimensional representation of the single pulses dataset. After marginalizing over the position parameter, every single pulse will be represented by an (energy, width, amplitude) point in this 3-dimensional space. The amplitude has arbitrary units, the width has units of phase bins, and the energy has units of the product of the former two features.
\label{fig:features_3d}}
\end{figure}

As a result, every single pulse will be represented by an (amplitude, width, position, energy) point in a 4-parameter space. We can visualize this space by marginalizing over one of the parameters and plotting the features in a 3-dimensional space, as presented in Fig.~\ref{fig:features_3d}. In this figure, we can appreciate the diversity in single-pulse morphology, with a vast majority of single pulses conglomerating in a low-energy, low-amplitude cluster at the bottom of the diagram. As expected, this distribution shows that most single pulses have low S/N and are dominated by noise. On the other hand, in the middle section of Fig.~\ref{fig:features_3d} and above the low-amplitude cluster we find a smaller number of single pulses with higher amplitudes; these represent the high-fluence data we will attempt to isolate in this analysis.

\subsection{Clustering Algorithms}\label{sec:clustering}

In order to assign the single pulse features (i.e., as shown marginalized in Fig.~\ref{fig:features_3d}) into the corresponding pulse states, we surveyed different automatized clustering algorithms. We opted for unsupervised classification methods to eliminate the need for a training dataset. However, due to the large sample size ($N \sim 219000$ single pulses per frequency band), clustering algorithms with high computational costs are impractical for this analysis. Instead, we surveyed low-computation cost (memory usage $\le \mathcal{O}(N^2)$) clustering algorithms from the \texttt{SciKit Learn} \citep{scikit-learn} library.

\subsubsection{K-Means}

The K-Means algorithm \citep{macqueen1967some} is a method of vector quantization that clusters data by trying to divide a set of $N$ samples into $k$ disjoint clusters, each described by the mean position $\mu_{j}$ of its samples $x_i$ in the features space. The means are commonly called the cluster “centroids”. This algorithm aims to find centroids that minimize the within-cluster sum of squares criterion (i.e., variance). The algorithm has three steps:

\begin{enumerate}
    \item Choose random centroids, with the most basic method being to choose samples from the dataset.
    \item After initialization, loop between two steps:
    \begin{enumerate}[label*=\arabic*.]
        \item Each sample is assigned to its nearest centroid using an Euclidean distance.
        \item New centroids are created by taking the mean value of all the samples assigned to each previous centroid.
    \end{enumerate}
Compute the difference between the old and the new centroids and repeat the last two steps until this difference is smaller than a threshold.
\end{enumerate}

This algorithm requires the number of clusters to be provided. Moreover, it assumes that clusters are convex and isotropic, and responds poorly to irregular clusters. However, for this work it is reasonable to assume that all high-fluence single pulses will be distributed in an approximately convex area near the high-amplitude, low-width top region of Fig.~\ref{fig:features_3d}.

While a higher number of clusters results in finer morphology resolution, it also reduces the number of samples per cluster and, therefore, the number of single pulses available to create an integrated pulse profile. Therefore, we surveyed different numbers of clusters, from $k=2$ to $17$. Moreover, given the high dimensionality of the problem, it is convenient to run the k-means algorithm several times with different centroid seeds. Therefore, we ran the algorithm 3 times for each value of $k$. The final result is the best output of the consecutive runs in terms of intra-cluster inertia.

\subsubsection{Mean Shift}

Mean shift clustering \citep{1000236} is a density-based clustering algorithm that can be summarized as follows:

\begin{enumerate}
    \item For each data point, calculate the mean of all points within a certain radius (estimated using nearest-neighbor analysis) centered at the data point.
    \item Shift the position of the data point to that of the mean.
    \item Repeat steps 1 and 2 until a convergence criterion has been met. In each iteration, each data point will move closer to the mode (i.e., the highest density of data points in the region), which is or will lead to the cluster center.
    \item Identify the cluster centroid candidates as the points that have not moved after convergence.
    \item Assign each data point to the closest cluster centroid.
\end{enumerate}

Mean shift is particularly useful for datasets where the clusters have arbitrary shapes and are not well separated by linear boundaries. Unlike the K-means algorithm, it is non-parametric and does not require specifying the number of clusters in advance, since this is determined by the algorithm with respect to the data. However, it is more computationally expensive ($\sim \mathcal{O}(N^2)$), and it is not guaranteed that the resulting number of clusters is optimal for a given dataset.

\subsubsection{DBSCAN}\label{subsec:dbscan}

The Density-Based Spatial Clustering of Applications with Noise \citep[DBSCAN,][]{Ester1996ADA, 10.1145/3068335} is an unsupervised clustering method based on a threshold for the number of neighbors, \texttt{min\_samples}, within the radius \texttt{eps} according to some metric (i.e., euclidean distance). A data point with more than \texttt{min\_samples} neighbors within this radius is considered a core point, and all those neighbors (called direct density reachable) are considered to be part of the same cluster as the core point. If any of these neighbors is again a core point, their neighborhoods are combined into the same cluster. Non-core points in this set are called border points, and points that are not density reachable from any core point are considered noise.

Unlike K-Means, DBSCAN does not require assumptions about the shape and convexity of the clusters, and it can discover clusters of arbitrary shapes and sizes, including noise points (outliers) in the data. However, because \texttt{eps} is fixed for all points, the algorithm struggles when clusters have significantly different densities.

Instead of specifying the number of clusters, DBSCAN requires values of \texttt{min\_samples} and \texttt{eps} to be provided. Besides some heuristic-based approaches \citep{1998DMKD....2..169S}, choosing appropriate values can pose a challenge. The choice depends on each dataset and experimentation with different values is usually needed to achieve the desired clustering results. To obtain results analogous to those obtained using K-Means, we varied $\texttt{eps}=0.51, 0.52, \ldots, 1.1$ using a Euclidean metric, and \texttt{min\_samples} to cover $1\%, 1.5\%, \dots, 5\%$ of the total number of single pulses; we then retained the first combination of these values that resulted in $k=2, 3, \ldots, 17$ clusters.

DBSCAN requires a function to calculate the distance between data points. However, for high-dimensional data, there is little difference in the distances between different pairs of points and the metric can be rendered almost useless due to the so-called ``curse of dimensionality''\citep[e.g.,][]{bellman2003dynamic}, making it difficult to find an appropriate value for \texttt{eps}. Finally, DBSCAN visits each point of the database, possibly multiple times. As a result, the worst-case memory complexity of DBSCAN is $\sim \mathcal{O}(N^2)$,

\subsubsection{OPTICS}\label{subsec:optics} 

The Ordering points to Identify the Clustering Structure (OPTICS) algorithm \citep{ankerst1999optics} can be seen as a generalization of DBSCAN that relaxes the \texttt{eps} parameter from a single value to a range bounded by \texttt{max\_eps}, which is the maximum radius from each point to find other potential reachable points. Each point is then assigned two distances:
\begin{itemize}
    \item A \textit{core distance}, given by the distance to the \texttt{min\_samples}-th closest point.
    \item A \textit{reachability distance}, given by the core distance or the distance to the nearest neighbor, whichever is bigger.
\end{itemize}

\noindent Clusters can then be extracted by selecting a threshold on the reachability distance, or by different algorithms that try to detect steepness in a \textit{reachability plot}, where the points are linearly ordered such that spatially closest points become neighbors in the ordering.

By using a range of radius values, OPTICS handles clusters with varying densities more effectively than DBSCAN and can identify noise points at multiple levels. However, noise points are not as well-defined as in DBSCAN because the first samples of each area have a large reachability and will thus sometimes be marked as noise. The choice of its two hyperparameters can also pose a challenge, and we resorted to proving the same ranges for \texttt{max\_eps} and \texttt{min\_samples} as those used for DBSCAN (see Sec.~\ref{subsec:dbscan}).

OPTICS clustering can be computationally expensive and slow for large datasets. It also requires more memory than DBSCAN, which can be a problem for datasets with limited memory \citep{Schubert2018ImprovingTC}.

\subsection{TOA Uncertainty Calculation}\label{subsec:toa_calculation}

\cite{2014MNRAS.441.3148O} sought to reduce the RMS timing residual by removing the single pulses with the lowest S/N from the integrated pulse profile average. Such a scheme can potentially yield an improved TOA uncertainty for short observations. However, when larger numbers of single pulses are available, we found that the improvement in TOA uncertainty resulting from averaging a sufficiently large number of single pulses outweighs that from averaging only the brightest pulses. Therefore, instead of discarding the lower-S/N data, we aim to gather the information from all the pulse states while weighting the contribution of each pulse according to its fluence state. To this end, we devised the following algorithm to compute a weighted $\mathrm{TOA}$ measurement:

\begin{enumerate}
    \item We used the single-pulse features obtained in Sec.~\ref{subsec:features} as the training instance for a clustering algorithm (see Sec.~\ref{sec:clustering}). As a result, every single pulse was classified into one of $k$ disjoint clusters, labeled $s_{0}, \ldots, s_{k-1}$.
    \item We calculated the integrated pulse profile of each cluster using the dataset that was not normalized by the dynamic spectrum.
    \item We used \texttt{PyPulse}'s \texttt{component\_fitting} function to fit multiple Gaussian components to this pulse profile \citep[e.g.:][]{Lam_2019}. The fit is performed iteratively until either adding more components is deemed insignificant via an F-test with a significance value of $\alpha=0.05$ (corresponding to a "$2\sigma$" level), or 10 components are fit. We thereby obtained a smoothed copy of the integrated pulse profile, which we adopted as the pulse template for that cluster. 
    \item We used \texttt{PyPulse}'s \texttt{fitPulse} function to match this template against the corresponding template profile using the phase gradient scheme \citep{1992RSPTA.341..117T} in the Fourier domain. As a result, we obtained the time-of-arrival ($t_{i}$) and TOA uncertainty ($\sigma_{i}$), corresponding to cluster $s_i$. 
    \item Each cluster's TOA measurement $t_i$ was assigned a weight given by inverse-variance weighting: $w_i=1/\sigma_{i}^2$.
    \item Finally, we calculated the inverse-variance weighted average of $\{t_{0}, \dots, t_{k-1}\}$ using  $\{w_0, \ldots, w_{k-1} \}$ as the corresponding weights, given by:

 \begin{equation}
     \bar{t} = \frac{\sum_{i=0}^{k-1} t_i w_i}{\sum_{i=0}^{k-1} w_i}
 \end{equation}

Since the $t_{i}$ measurements are all independent, the variance of the weighted mean is given by \citep[e.g.:][]{weighted_variance, hartung2011statistical}:

\begin{equation}
    \sigma_\mathrm{TOA}^2 = \frac{\sum_{i=0}^{k-1} w_i^2 \sigma_i^2}{\left( \sum_{i=0}^{k-1}{w_i} \right)^2} = \frac{1}{\sum_{i=0}^{k-1} 1/\sigma_i^{2}} 
\end{equation}

\end{enumerate}

By using the weighted mean of the $\mathrm{TOA}$s from all clusters, as opposed to removing the lowest S/N pulses like in \citep{2014MNRAS.441.3148O}, we ensure that all the available data is utilized in the $\mathrm{TOA}$ calculation. Moreover, by weighting each pulse state by the corresponding $\sigma_i$, as opposed to weighting all the states equally like in traditional timing techniques, the high-fluence pulse states contribute more prominently to the $\mathrm{TOA}$ calculation than low-fluence states.

\subsection{RFI Mitigation}\label{subsec:RFIs}

Radio frequency interference (RFI) can severely hinder the timing sensitivity of even the most sophisticated radio telescopes. Conventional pulsar timing techniques can reduce the effects of transitory RFI by folding large numbers of single pulses. However, since the present analysis requires clustering the data into smaller subsets of single pulses, the effects of unfiltered RFI will be more prominent in the resulting cluster average pulse profile, therefore greatly affecting the timing precision. As a result, our method is highly susceptible to RFIs. In particular, by visually inspecting pulse profile samples we found that the L-band data was more heavily affected by RFIs compared to the 820-band data, so employing RFI-excision techniques was paramount to this analysis. To such end, we evaluated the merit of five RFI removal algorithms:

\begin{itemize}
    \item \texttt{MeerGuard} \citep{2020ascl.soft03008L} uses a frequency-dependent template to identify an off-pulse region and calculate the statistical features of that region (mean, standard deviation, peak-to-peak amplitude, etc). The features are compared to those of other pulses in the same frequency channel and the same single pulse sub-integration, and outliers are flagged for RFI removal.
    \item \texttt{clfd} \citep{2019MNRAS.483.3673M} assigns every single pulse to a set of up to three numerical profile features: its standard deviation, peak-to-peak difference, and the absolute value of the second frequency that results from discretizing the Fast Fourier Transform of the profile. Outliers are then flagged in the resulting feature space. To avoid filtering bright single pulses, we excluded the peak-to-peak difference in the analysis.
    \item \texttt{zap\_minmax} is NANOGrav's median smoothed difference channel zapping algorithm, built on \texttt{PSRCHIVE}. A median smoothing window 21-frequency channels wide is applied to the total variation of the off-pulse noise, and all frequency channels with total flux greater than 4 standard deviations away from the median in that window are flagged for removal.
    \item \texttt{RFIClean} \citep{2021A&A...650A..80M} performs Fourier-domain excision of periodic RFI, and threshold-based techniques to identify broadband bursts and narrowband RFI. We first converted our observations from PSRFITS to filterbank format using the \texttt{dspsr} package, cleaned using \texttt{RFIClean}, and converted back to PSRFITS using the \texttt{your} \citep{Aggarwal2020} package.
    \item \texttt{paz}, from the software suite \texttt{PSRCHIVE}, combines median smoothed difference channel zapping (\texttt{paz -r})  with an algorithm that replaces phase bins of abnormally high intensities with the local median plus noise (\texttt{paz -L}).
\end{itemize}

We find that the general conclusion of improvement via clustering does not depend on the choice of RFI excision algorithm, only the resulting TOA uncertainties overall. The results of this analysis are presented in Sec.~\ref{sec:rfi}.

\section{Results}\label{sec:results}

Here we present the TOA uncertainties obtained when clustering schemes are applied, we analyze the structure of the corresponding clusters in feature space, and we discuss how the results vary in the presence of different noise levels.

\subsection{Impact of RFI Excision Algorithm}
\label{sec:rfi}

\begin{figure}[t]
\hspace*{-0.7cm}
\epsscale{1.3}  
\plotone{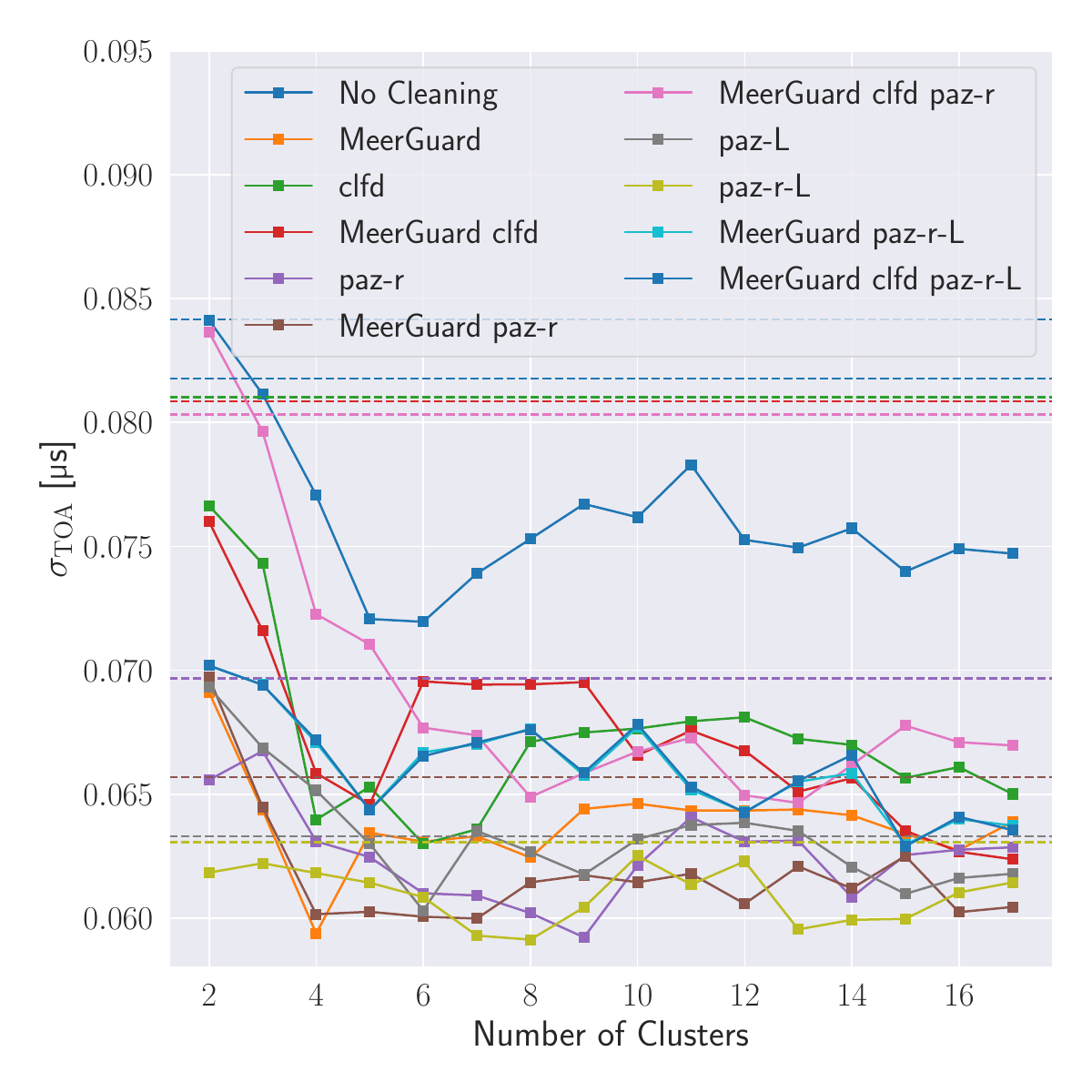}
\caption{TOA errors obtained using different combinations of the RFI-excision algorithms presented in Sec.~\ref{subsec:RFIs} to process the 820-band dataset before applying a K-means clustering algorithm. The dashed horizontal lines represent the TOA error obtained for each algorithm when no clustering is applied. \label{fig:recipes}}
\end{figure}

To assess the most efficient RFI-excision recipe and the impact on our clustering method, we implemented different combinations of the previously described RFI-excision algorithms (see Sec.~\ref{subsec:RFIs}) on the 820-band data and measured the resulting $\sigma_\mathrm{TOA}$ as a function of the number of clusters when using the K-means clustering algorithm. The results are summarized in Fig.~\ref{fig:recipes}. We find that the resultant TOA uncertainty varies greatly depending on the RFI-excision algorithm, so we can expect that the efficiency of our timing method will be highly dependent on the RFI content of the observation. However, we see broadly that multiple clusters improve timing precision regardless of the algorithm used. In the next subsections, we opted for the RFI-excision algorithm that provided the lowest $\sigma_\mathrm{TOA}$ median; for the 820-band dataset this is a combination of \texttt{MeerGuard} and \texttt{paz -r}, and for the more heavily RFI-affected L-band it is a combination of \texttt{MeerGuard}, \texttt{clfd}, and \texttt{paz -r}. 

\subsection{TOA Uncertainties}

\begin{figure}[t]
\hspace*{-0.7cm}
\centering
\epsscale{1.3}
\plotone{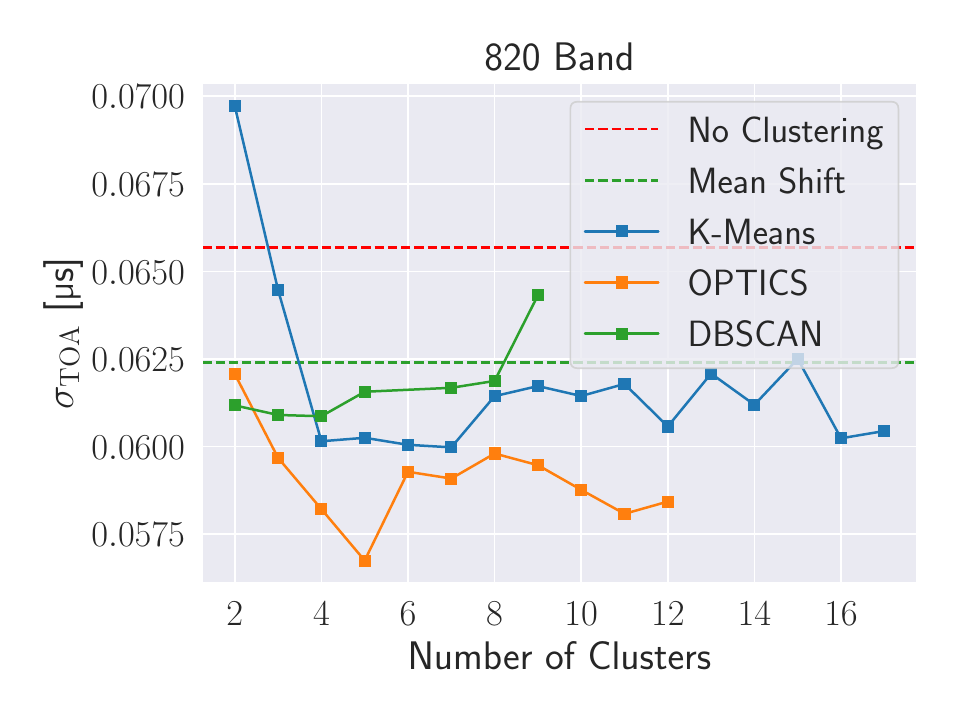}
\caption{Weighted average TOA errors obtained by each clustering algorithm (see Sec.~\ref{sec:clustering}) applied to the 820-band data, as a function of the number of clusters found by the algorithm. The horizontal dot-dashed red line corresponds to the TOA error obtained when no prior clustering of the single pulses is applied, and we integrate over the entire observation and frequency band.}
\label{fig:clustering_results_820}
\end{figure}

The TOA uncertainties obtained when clustering the 820-band data using each clustering algorithm are discussed in Fig.~\ref{fig:clustering_results_820}. In particular:

\begin{itemize}
    \item When no prior clustering of the single pulses is applied and we average the data in time and frequency, for this dataset we obtained a TOA uncertainty of $\sigma_\mathrm{TOA}^{(0)}=0.066~\mu \mathrm{s}$.
    \item For the K-means algorithm, the number of clusters was varied from $k=2$ to $17$. For $k\geq 3$ we obtained a systematic improvement in $\sigma_\mathrm{TOA}$. In particular, we attained the lowest TOA uncertainty at $k=7$ clusters resulting in $\sigma_\mathrm{TOA}=0.06~\mu \mathrm{s}$, which represents a reduction of $\sim 6~ \mathrm{ns}$ (relative ratio of $\Delta \sigma_\mathrm{TOA} / \sigma_\mathrm{TOA}^{(0)} = 0.087$). Sub-optimal results are obtained when the number of clusters is too small ($k \sim 2$). 
    \item For Mean Shift, the number of clusters was determined automatically by the algorithm to be $18$ and we obtained $\sigma_\mathrm{TOA}=0.062~\mu \mathrm{s}$ ($\Delta \sigma_\mathrm{TOA} / \sigma_\mathrm{TOA}^{(0)} = 0.049$); while this is an improvement over the non-clustering error, it is usually outperformed by the K-Means, arguably because the number of clusters found by Mean Shift is not optimal for this dataset. 
    \item For DBSCAN and OPTICS, we surveyed different combinations of its two hyperparameters (see Sec.~\ref{sec:clustering}), resulting in a more limited range of clusters. We found that DBSCAN attains its best results at lower numbers of clusters ($k \leq 4$); otherwise, its performance is comparable to that of K-Means and Mean Shift. Conversely, OPTICS performs systematically better than K-Means and attains its lowest TOA uncertainty of $\sigma_\mathrm{TOA}=0.057~\mu$s at $\mathrm{min\_samples}=2136$ (1\% of the total number of single pulses) and \texttt{eps}=0.77. This is a reduction of $\sim 9~\mathrm{ns}$ ($\Delta \sigma_\mathrm{TOA} / \sigma_\mathrm{TOA}^{(0)} = 0.136$) compared to uncertainty when no clustering is applied.
\end{itemize}

\begin{figure*}[htp]

\centering
\subfigure[]{\includegraphics[width=8.75cm]{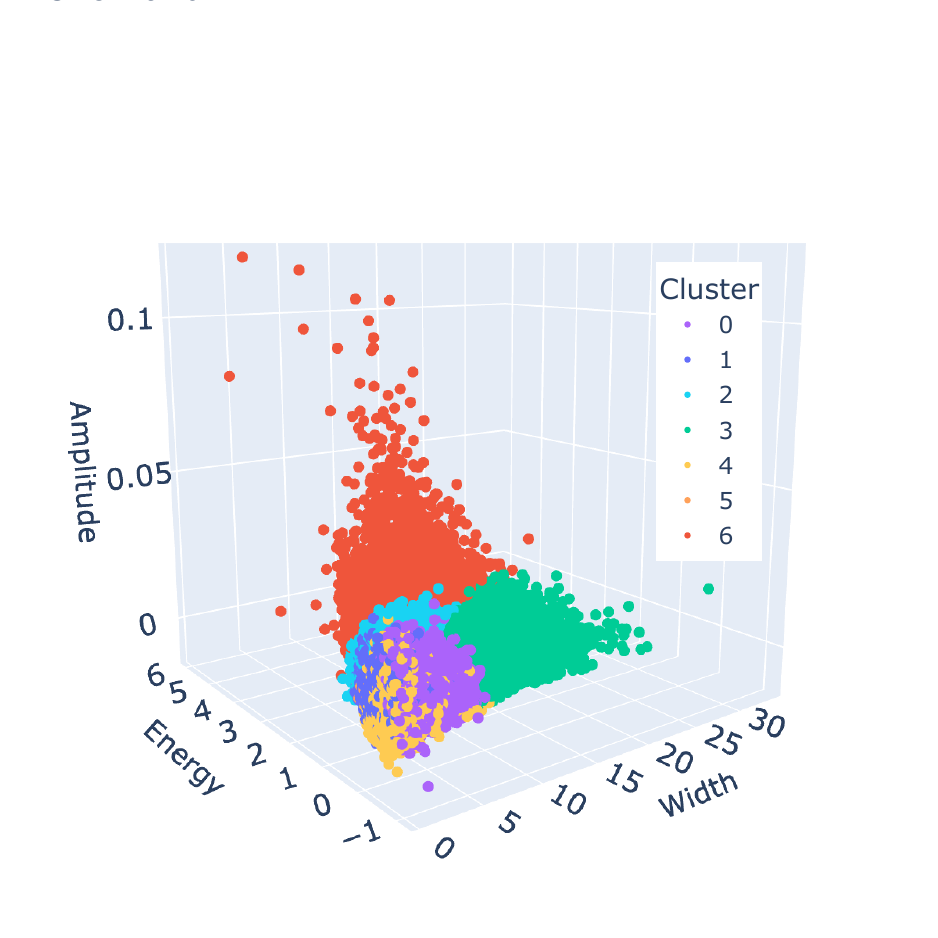}\label{fig:clustered_3d_820}}
\centering
\subfigure[]{\includegraphics[width=8.75cm]{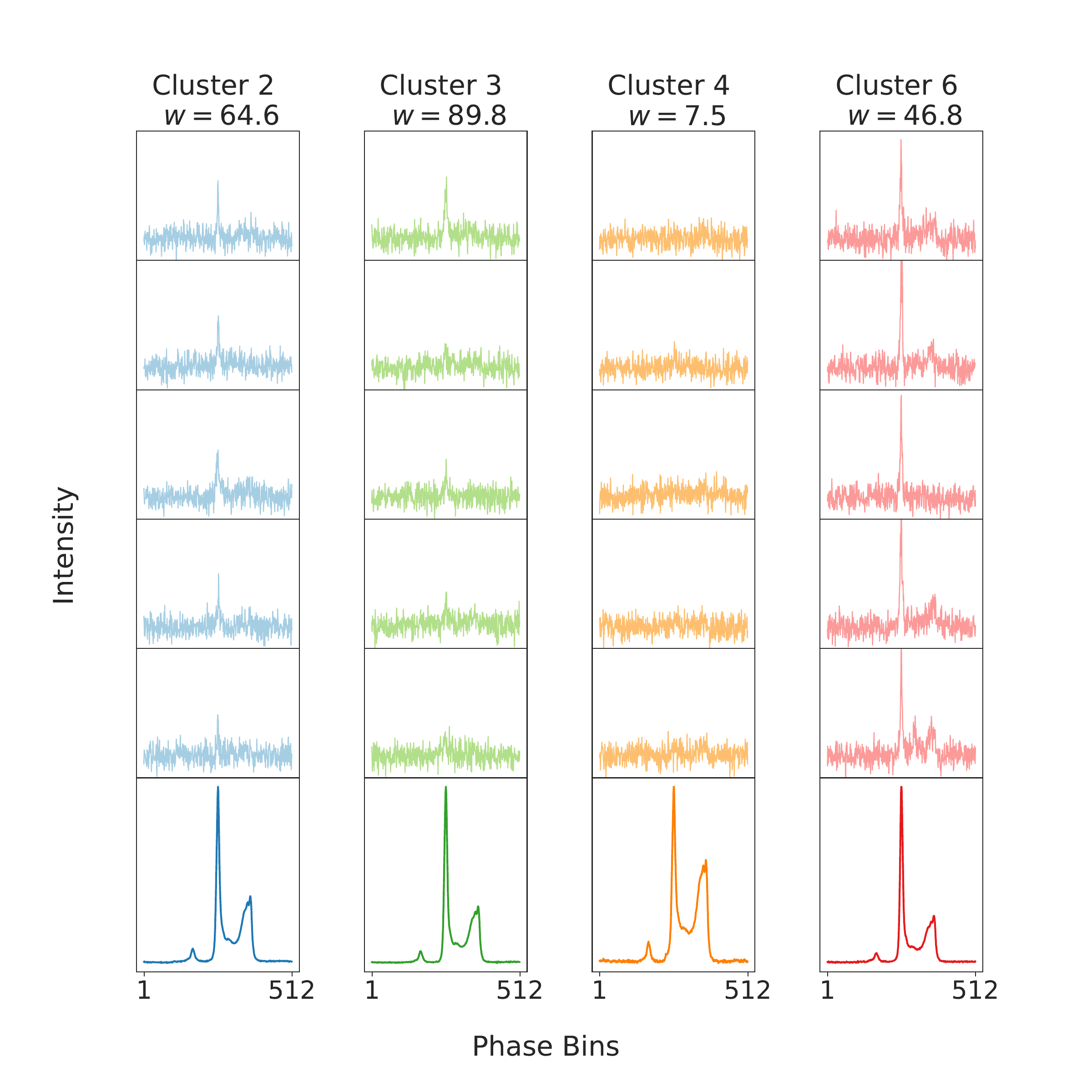}\label{fig:clusters_examples}}

\caption{Left: 3-dimensional representation of clusters obtained using the K-means algorithm with $k=7$ on the 820-band single pulses. Right: examples of single pulses classified into clusters. Each column corresponds to a different cluster, and the same color scheme for each cluster was used across both plots. At the top of each column, $w_i=1/\sigma_{i}^2$ indicates the weight assigned to each cluster (see Sec.~\ref{subsec:toa_calculation}). In the last row of each column, we present the integrated pulse profile corresponding to that cluster.
\label{fig:parameter_fit}}\label{fig:3d_clusters_820_band}

\end{figure*}

\subsection{Clustering Structure}

Both K-Means and OPTICS provide the most optimal results; however, since K-Means only requires one hyperparameter to be specified, its implementation is significantly more straightforward. Therefore, we used it as a case study to examine its underlying clustering structure. In Fig.~\ref{fig:clustered_3d_820} we present the classified data in a 3-dimensional features space when using $k=7$ clusters. In Fig.~\ref{fig:clusters_examples} we present some of the single pulses assigned to four of these clusters. We observe that each cluster corresponds to a distinctive single pulse morphology:
\begin{itemize}
    \item Clusters 4 and 5 conglomerate towards low amplitudes and are therefore composed of low-S/N pulses dominated by background noise. While they also show low widths, this can be attributed to the low energies of these single pulses, as evidenced in Fig.~\ref{fig:clusters_examples}.
    \item Clusters 2 and 3 present higher amplitude single pulses, but those in cluster 3 are wider than in cluster 2. 
    \item Cluster 6 leans towards the high-amplitude, low-width, high-energy part of the features space and is thereby comprised of bright, high-S/N single pulses.
\end{itemize}

\noindent The last row of Fig.~\ref{fig:clusters_examples} shows the integrated pulse profile corresponding to each cluster. We readily observe that single pulses in different clusters average to different shapes, with varying peak heights, widths, and relative amplitudes between the pulse components. This result agrees with \cite{2014MNRAS.443.1463S}, where the authors observed a similar behavior when using the single pulses with the highest intrinsic energy, and extends those findings to single pulses in other pulse states.

\begin{figure}[h]
\epsscale{1.2}  
\plotone{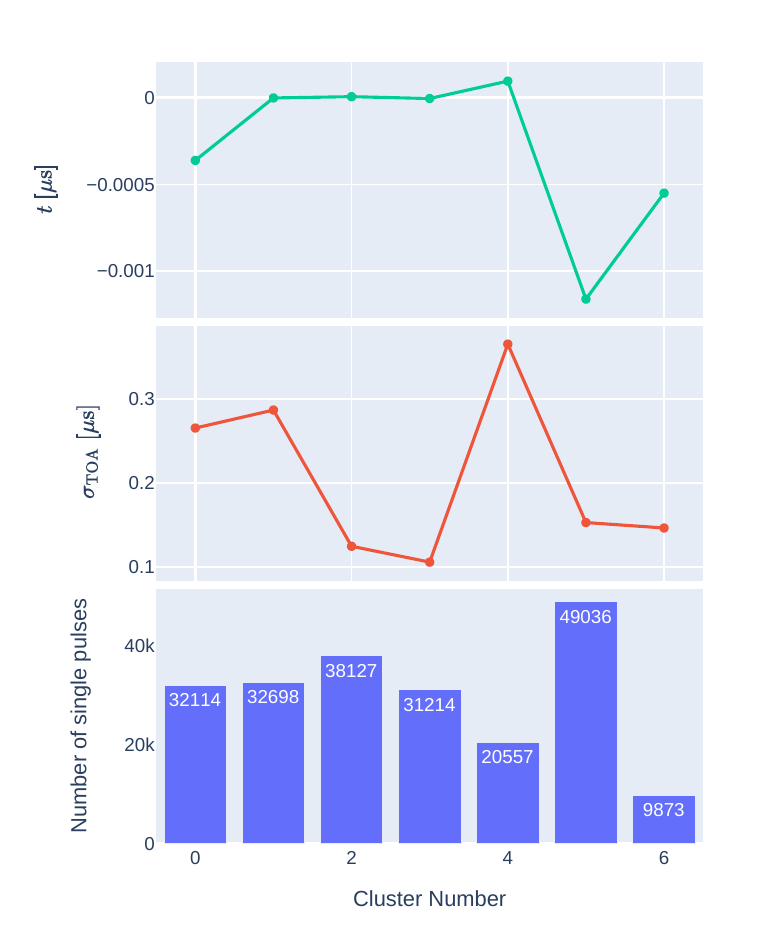}
\caption{$t$, $\sigma_\mathrm{TOA}$, and the number of single pulses for each of the clusters found in the 820-band dataset using a K-Means algorithm with $k=7$ clusters. 
\label{fig:7_clusters_results}}
\end{figure}

In Fig.~\ref{fig:7_clusters_results} we show the $\mathrm{TOA}$ and $\sigma_\mathrm{TOA}$ resulting from each cluster, as well as the number of single pulses per cluster. We find that clusters 2, 3, 4 present reduced TOA uncertainties due to averaging a large number of mixed-fluence single pulses, which is the conventional procedure in pulsar timing. However, cluster 6 also attains a comparably low TOA uncertainty despite comprising the smallest number of single pulses. Indeed, this cluster represents a high-fluence state, and the reduction in $W_\mathrm{eff}$ due to only averaging high-S/N single pulses is enough to outweigh the improvement in $S$ that would result from averaging a large number of single pulses, resulting in a smaller $\sigma_\mathrm{TOA}$ (see Eq.~\ref{eq:new_sigma_TOA}). Conversely, cluster 4 does not benefit from a large number of single pulses or being a high-fluence state, resulting in a large TOA uncertainty.

\begin{figure}[h]
\hspace*{-0.7cm}
\centering
\epsscale{1.29}
\plotone{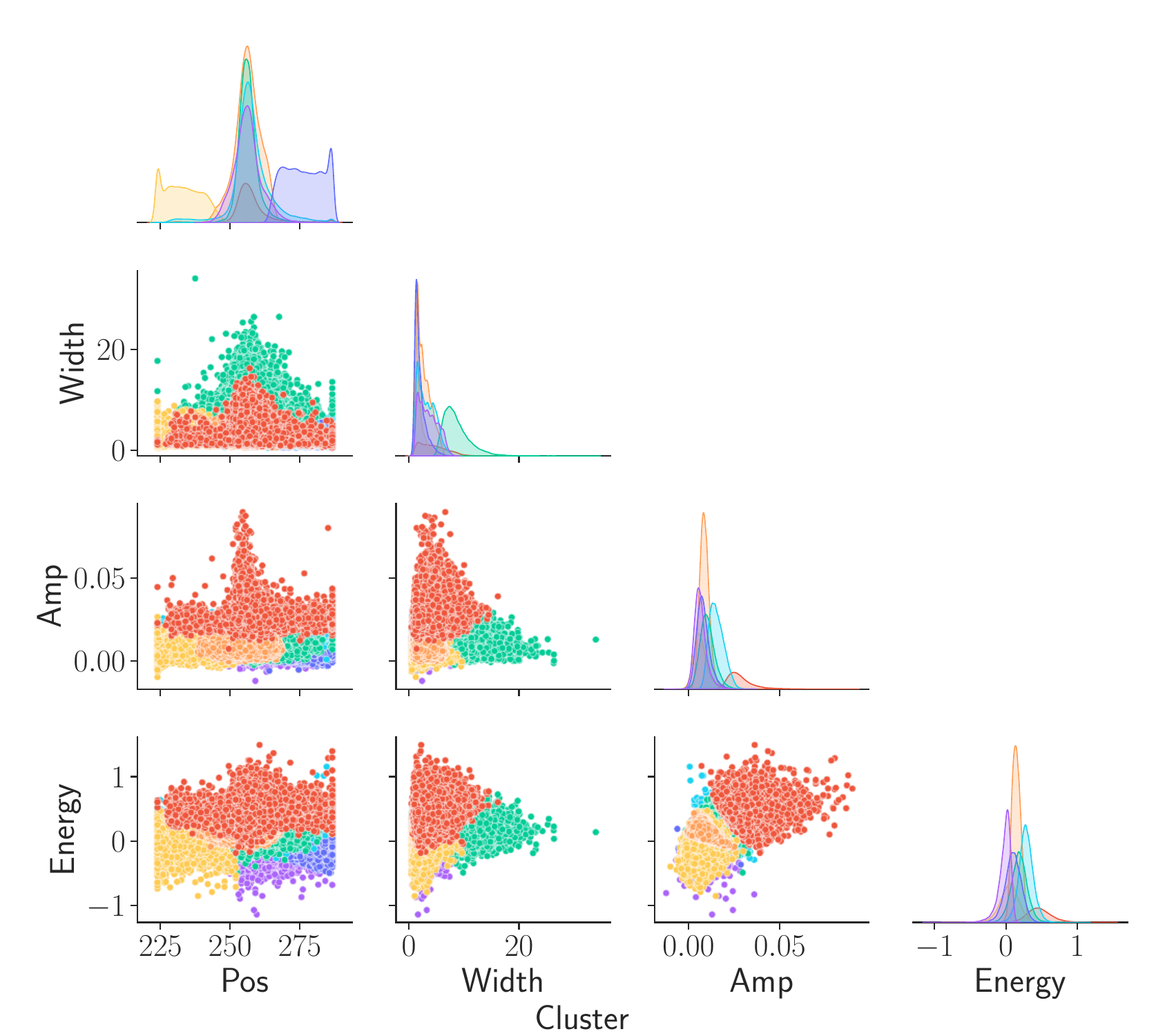}
\caption{Pairs plot for the single pulses in the 820-band data clustered using K-means with $k=7$ clusters. These show the relationships in the distributions of the 4 single pulse features: pulse position, width, amplitude, and energy. In the main diagonal, kernel density estimate plots show the marginal distribution of the data in each feature. The color scheme used to identify the different clusters matches that in Fig.~\ref{fig:3d_clusters_820_band}.
\label{fig:pairplots}}
\end{figure}

In Fig.~\ref{fig:pairplots} we present a pairs plot showing the distribution of the clustered single pulses in single-pulse features space. By looking at the distribution in pulse position, we see that the clustering algorithm successfully identifies pulses that fall on the edges of the main pulse window, possibly because no significant pulse peak can be identified (clusters 4 and 1). We also note a clear separation in amplitude (with cluster 7 skewed to the highest amplitudes and energies) and width (cluster 3 skewed to the highest widths).

\begin{figure}[h]
\epsscale{1.2}  
\plotone{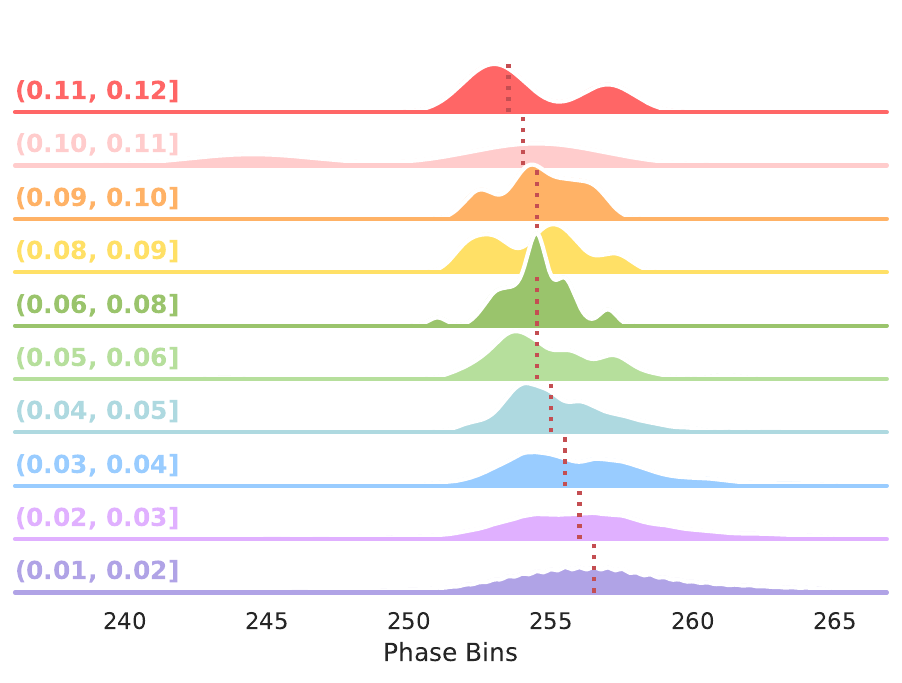}
\caption{Distribution of the pulse position (in phase bins units) for the single pulses in the 820-band dataset, after separating them in intervals of pulse amplitudes. Each row shows the distribution for a given amplitude interval; the vertical red dotted line represents the median of the pulse positions in that interval. The amplitudes are in arbitrary units because the observations are not flux-calibrated.
\label{fig:ridgeplot}}
\end{figure}

In analyzing Fig.~\ref{fig:pairplots}, we noticed a trend for peaks with larger amplitudes to appear thinner and have earlier positions than pulses with lower amplitudes. In particular, the distribution in peak location of the high-fluence cluster 6 has a mode towards earlier phase bins than low-fluence cluster 5; intermediate-fluence clusters 2 and 3 fall in the middle. To better study this behavior, we grouped the single pulses in intervals of increasing amplitude and, for each interval, we calculated the distribution across the position of the main pulse peak. The results are presented in Fig.~\ref{fig:ridgeplot}. We find that, on average, a higher amplitude is correlated with a more leftward position of the main pulse and, therefore, an earlier time of arrival of the main pulse component. By calculating the median of the positions of the single pulses in each interval (vertical red dotted lines in Fig.~\ref{fig:ridgeplot}), we find that the pulses with the highest amplitudes arrive 3 phase bins earlier than the lowest-amplitude ones, which corresponds to a time offset of $94.04~\mu$s.

\subsection{L-band and Injected Noise}

Finally, we performed a similar analysis for the single pulses in the L-band dataset. Due to the lower S/N for the observation in this band, the resulting distribution across the 4 single-pulse features is more narrow than for the 820-band data, with a predominance of low-amplitude, low-energy single pulses, which corresponds to a majority of single pulses dominated by noise. As a result, the clustering algorithms resulted in mixed results in determining meaningful clusters. Moreover, a wider exploration of the hyperparameter space was needed to initialize the algorithms correctly. We also noted that the high-fluence clusters are less populated in the L-band data and potentially do not meet the number of single pulses required to obtain a stable integrated pulse shape, resulting in a decreased TOA precision.

\begin{figure}[t]
\centering
\epsscale{1.2}
\plotone{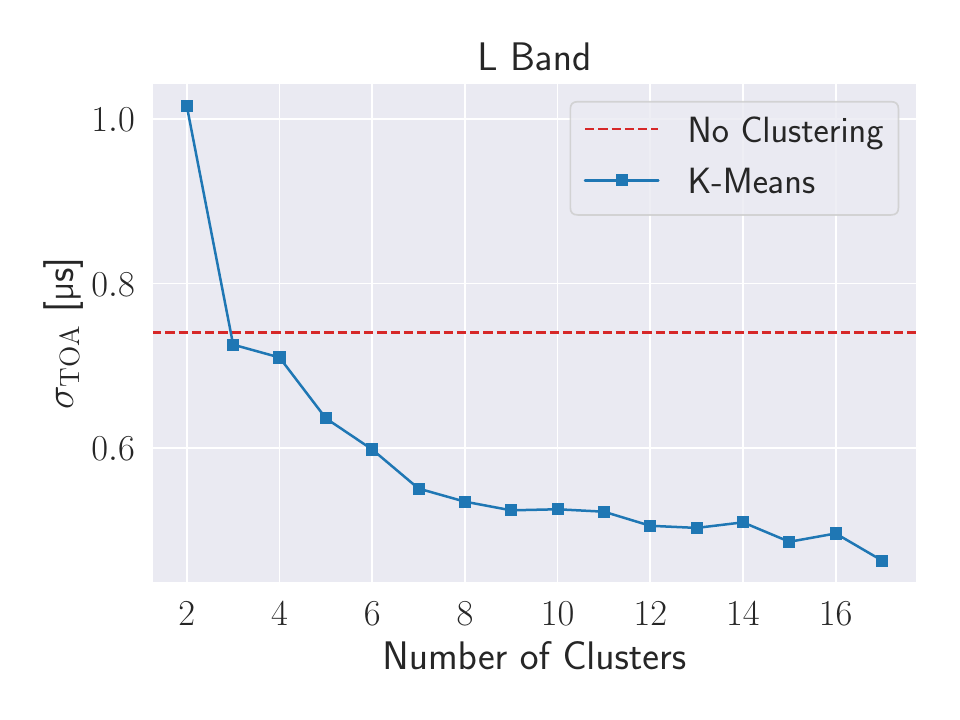}
\caption{Weighted average TOA errors obtained by K-Means applied to the L-band data, as a function of the number of clusters found by the algorithm. The horizontal dot-dashed red line corresponds to the TOA error obtained when no prior clustering of the single pulses is applied.}
\label{fig:clustering_results_Lband}
\end{figure}

The most robust results were found using a K-Means algorithm, presented in Fig.~\ref{fig:clustering_results_Lband}. When no clustering scheme is used and all single pulses are weighted equally, we obtained a TOA error $\sigma_\mathrm{TOA}^{(0)}=0.74~\mu$s. When a K-Means classifier was applied to weigh the data, we obtained TOA errors as small as $\sigma_\mathrm{TOA}=0.46~\mu$s, which represents an improvement of $0.28~\mu$s ($\delta \sigma_\mathrm{TOA} = 0.37$).

\begin{figure}[t]
\epsscale{1.2}  
\plotone{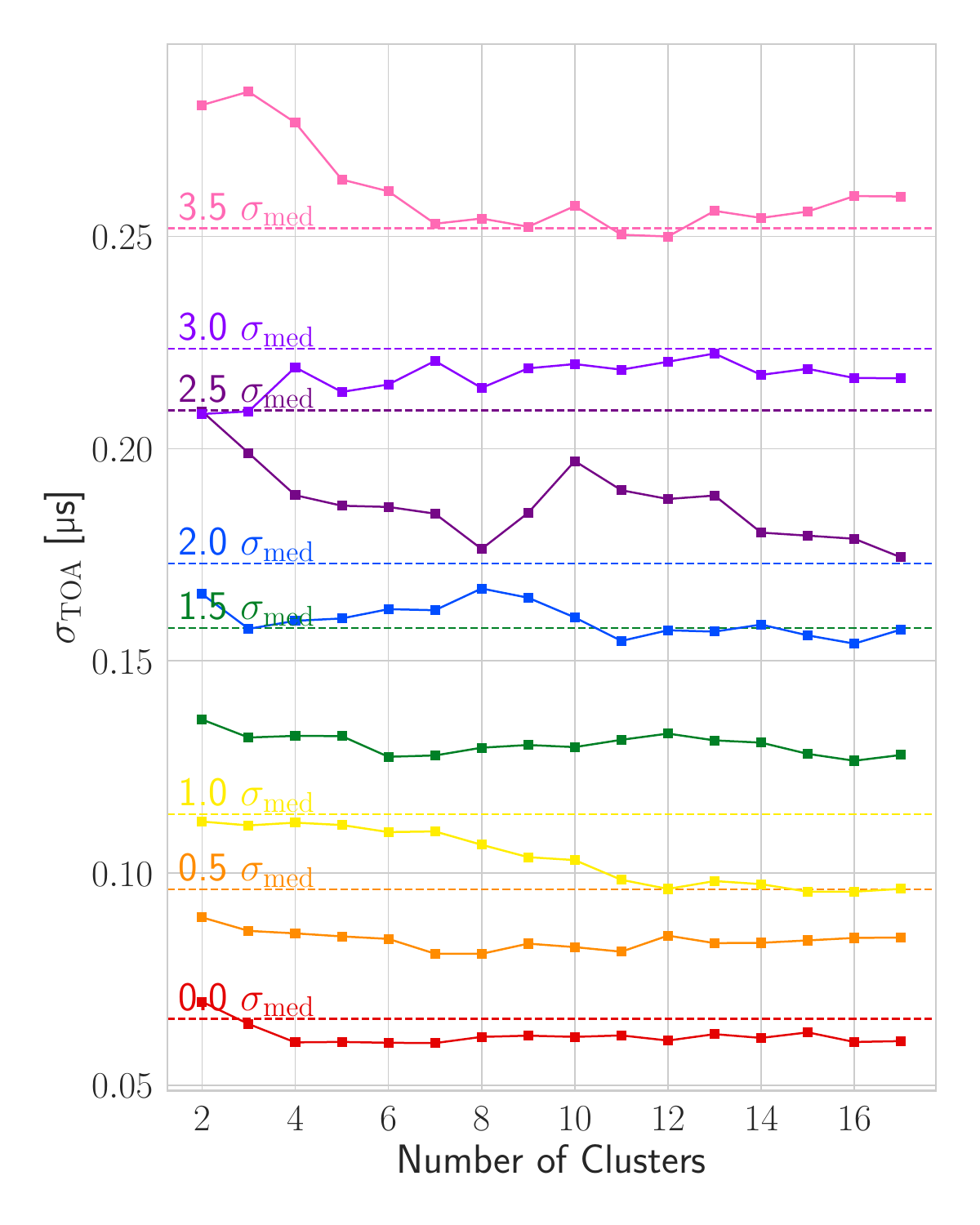}
\caption{$\sigma_\mathrm{TOA}$ as a function of the number of K-means clusters when adding different noise levels to the 820-band dataset. Each level of noise is presented with a different color. The dashed lines represent the $\sigma_\mathrm{TOA}$ achieved without clustering techniques. The dot-solid lines represent the results when using K-means for a given number of clusters.
\label{fig:noise_levels}}
\end{figure}

The clustering algorithms we tested were more successful at correctly identifying meaningful fluence states for the 820-band data than for the L-band data. This behavior can be attributed to the difference in S/N between both frequency bands. Therefore, to quantify the robustness of this method when applied to datasets of varying S/N, we repeated the analysis on observations with injected artificial noise. For every single pulse in the 820-band dataset, we calculated the root-mean-square of the off-pulse intensities, $\sigma_\mathrm{OPI}$, and then added white noise with a standard deviation equal to $0.5~\sigma_\mathrm{OPI}$, $1.0~\sigma_\mathrm{OPI}$, $1.5~\sigma_\mathrm{OPI}$, etc. 
For each level of injected noise, we classified the resulting data using K-means and calculated the weighted weight-averaged $\sigma_\mathrm{TOA}$ using the algorithm described in Sec.~\ref{sec:data_analysis}; the results are presented in Fig.~\ref{fig:noise_levels}. We find that for injected noise levels up to $3.0~ \sigma_\mathrm{OPI}$, our method provides an improvement in $\sigma_\mathrm{TOA}$ over conventional techniques with no clustering. However, for an injected noise amplitude of $3.5~ \sigma_\mathrm{OPI}$ and higher, the conventional method outperforms our clustering approach.

\section{Conclusions}\label{sec:conclusions}

In this work, we proposed that the stochastic variations in single-pulse morphology correspond to shifts between a range of pulse fluence states. Moreover, we created an algorithm for classifying single pulses according to their fluence to weigh their contribution to the TOA measurement for the epoch. We then tested the potential of this method to decrease the uncertainty in the TOA measurement by testing it on observations of PSR~J2145$-$0750. The algorithm performance depended on the observation, the choice of the clustering algorithm, and the corresponding hyperparameters. 
For an observation in the $820$-MHz frequency band, we found that both K-Means and OPTICS provide similar improvements in the TOA uncertainty calculated ($\sigma_\mathrm{TOA}=0.06~\mu$s and $0.057~\mu$s, respectively) compared to conventional timing techniques with no clustering ($\sigma_\mathrm{TOA}=0.066~\mu$s). However, OPTICS requires a length hyperparameter tunning and is considerably slower and less robust when applied to high-dimensional data (see Sec~\ref{subsec:optics}). As a result, we find that OPTICS can be computationally expensive and difficult to initiate for single pulse datasets. Conversely, K-Means is simple to implement, for it only requires one hyperparameter, and works well on larger datasets. Therefore, OPTICS can be optimized for more limited datasets, but K-Means can provide more robust and computationally efficient performances for larger datasets consisting of several observations and different pulsars.

In this analysis, we used the weighted average TOA and TOA error as a metric to quantify the timing precision of our method. As a consequence of using different templates to calculate a TOA for each fluence cluster, the resulting TOA will be shifted relative to the one obtained from averaging all pulses with no intermediate clustering. Such shifts are only of consequence in terms of TOA accuracy whereas, for this work, we are only concerned about changes in TOA precision. However, since all single pulses belong to the same observations and are referenced to the same phase, we can potentially estimate the relative shift among the different clusters with this weighting.

In analyzing the distributions of the single pulse features, we found a tendency for higher-amplitude single pulses to arrive earlier in the pulse window and have a reduced pulse width compared to pulses with lower amplitudes. A correlation between pulse latitude and intensity was first observed in observations of the Vela pulsar (PSR B0833$-$45/J0835$-$4510) by \cite{1983ApJ...265..372K} and then again in PSR B$0329$+$54$ by \cite{1993A&A...269..325M} using the VLA observatory. The latter interpreted this behavior in terms of \cite{1980ApJ...235..576C}'s ion outflow model as a change in the emission heights of the individual pulse components. A similar correlation was observed by \cite{2012ApJ...761...64S} in PSR J1713+0747 using the Arecibo Telescope, and then again by \cite{2015MNRAS.452..607K} in PSR J0835$-$4510 (Vela) using the Parkes telescope. \cite{2022MNRAS.509.5790L} and \cite{2023MNRAS.521.4504Z} also found that pulses with larger amplitudes appear earlier than pulses with lower amplitudes in observations of the Vela pulsar at the Argentine Institute of Radioastronomy \citep{2020A&A...633A..84G}; they used this behavior to support a pulsar model based on emission regions at different altitudes in the neutron star magnetosphere where the pulses of each cluster are emitted. The discovery of the same correlation in independent observations across various pulsars and using different backends suggests that this is not an instrumental artifact or an isolated pulsar behavior, but rather a robust trend. This relative shift in phase could be quantified and incorporated into TOA calculations for improved accuracy; however, each cluster was assigned a pulse template consistent with this shift, so it should be of no consequence to the TOA precision. A better understanding of this phenomenon could be obtained by performing single-pulses analyses in observations of other pulsars. Future work could also explore the evolution of the position of the minor pulse components as a function of the fluence and quantify the heights of the emitting regions. Such a study could provide new insights into sub-pulse structure.


Among the limitations of this method, we recall that the dataset analyzed in this work benefited from the pulsar intrinsic high fluence, distinct pulse shape, and brightness magnification due to scintles in this particular observation (see Fig.~\ref{fig:dynamic_spectrum}). However, the L-band data presents noticeably higher levels of noise in the single pulses and a more narrow distribution across pulse features such as amplitude and width. Consequently, the clustering algorithms' ability to detect meaningful clusters was severely hindered when applied to this dataset. In that case, the $\sigma_\mathrm{TOA}$ was reduced from $0.74~\mu$s when no intermediate clustering was applied, to $0.46~\mu$s. This behavior is also observed when different noise levels were artificially introduced to the 820-band dataset. In particular, we found that when the root-mean-square of the off-pulse noise in the 820-band dataset was amplified by a factor of 3.5 from its original value, clustering the data did not significantly improve the TOA uncertainty.

We also found that the ability of the clustering methods to classify the single pulses into distinct fluence states is highly dependent on the RFI content of the observation. Indeed, unfiltered RFI can significantly change the single-pulse morphology, making low-S/N single pulses mistakenly identified as high-fluence due to the high amplitudes of the interference and, thereby, hindering classification efficiency. For this observation of J2145$-$0750, we found that a combination \texttt{MeerGuard} and \texttt{paz} provides the most adequate RFI filtering for the 820-band. However, the choice of RFI cleaning recipe will depend on the observing backend, frequency band, and RFI environment. As a result, we expect that this method is better suited for analysis involving high-S/N observations and bright pulsars, and only after careful RFI excision and sensible clustering hyperparameter tuning.

Another shortcoming is the large memory space needed to store single pulse datasets. A possible alternative to post-processing clustering is to develop a real-time clustering system. This would involve finding the optimal clustering and RFI-filtering scheme for a given pulsar ahead of the observing run. During the data acquisition process, subsets of the observed single pulses can be RFI-excised, classified in real-time into pre-established clusters, and then averaged into the corresponding cluster pulse profile. Once the observation is complete, the different cluster pulse averages can be timed separately and the TOAs from each cluster can be weighted average to obtain a single TOA. In doing so, we eliminate the need to store individual single pulse data while also preserving the underlying fluence structure.

This study provides a proof-of-principle technique to be applied more extensively to observations of other pulsars. Potential applications include real-time clustering of single pulses in observations of new bright pulsars to improve the precision of the TOA estimations. With increasingly sensitive telescopes, this method could be applied to long-term monitoring of individual pulsars to improve the TOA precision at each epoch and gain sensitivity in various tests of fundamental physics requiring precision timing experiments.

\begin{acknowledgments}
\textit{Acknowledgments and author contributions}: S.V.S.F. undertook the analysis, developed the code pipeline, and prepared the ﬁgures, tables, and the majority of the text. M.T.L. developed the mathematical framework for this work, selected the analyzed data set, assisted with the preparation of the manuscript, wrote the appendix, provided advice interpreting the results, and supervised the project development. M.A.M. and M.T.L. designed the observational setup and undertook the data acquisition at GBO.

We thank David Nice, Matthew Kerr, Yogesh Maan, and NANOGrav's Noise Budget and Timing Working Groups for their valuable input and feedback on the manuscript. We also thank Capella and Rosie for their canine support.

SVSF is supported by the National Science Foundation Graduate Research Fellowship Program under Grant No. 2139292. We acknowledge support from the NSF Physics Frontiers Center award number 2020265, which supports the NANOGrav project. The Green Bank Observatory and National Radio Astronomy Observatory are facilities of the NSF operated under a cooperative agreement by Associated Universities, Inc. MTL also acknowledges support from NSF AAG award number 2009468. S.V.S.F. acknowledges partial support from the NASA New York Space Grant, and from the Out to Innovate Career Development Fellowship for Trans and Non-binary People in STEM 2023. 

This work made use of Astropy:\footnote{http://www.astropy.org} a community-developed core Python package and an ecosystem of tools and resources for astronomy \citep{astropy:2013, astropy:2018, astropy:2022}. 

\end{acknowledgments}

%

\vspace{5mm}
\facilities{Green Bank Observatory (GBO).}


\software{PyPulse \citep{2017ascl.soft06011L}, PINT \citep{PINT}, clfd \citep{2019MNRAS.483.3673M}, MeerGuard \citep{2020ascl.soft03008L}, LMFIT \citep{LMFIT}, Astropy \citep{astropy:2013, astropy:2018, astropy:2022}, \texttt{dspsr} \citep{2011PASA...28....1V},  \texttt{your} \citep{Aggarwal2020}, \texttt{RFIClean} \citep{2021A&A...650A..80M}.}



\appendix

\section{Timing Precision Formalism for Two States}
\label{sec:appendix}

Here we consider the impact of two fluence states, one low and one high fluence state, in timing precision. We start with the RMS timing error, calculated from the template and the S/N as \citep{1983ApJS...53..169D}
\begin{equation}
    \sigma_{\mathrm{S/N}} = \frac{ \left[\displaystyle \int \displaystyle \int dt~dt'~\rho (t-t') U'(t) U'(t') \right]^{1/2} }{ S \displaystyle \int dt \left[ U'(t) \right]^{2} },
\end{equation}

\noindent where $\rho(\tau = t - t') \equiv \langle n(t) n(t+\tau) \rangle / \sigma_{n}^{2}$ is the autocovariance function of the noise. Following Eq.~\ref{eq:temp}, again $U(t)$ is the normalized (to unit height) pulse template shape, $n(t)$ is additive noise with RMS amplitude $\sigma_{n}$, and the S/N of the pulse profile is $S$. When the noise is uncorrelated and the pulse profile is sampled at intervals $\Delta t$ much shorter than the pulse width, we have $\rho (t) = \delta(\tau) \Delta t$, which means the TOA error reduces to

\begin{equation}
    \sigma_{\mathrm{S/N}} = \frac{\Delta t^{1/2}}{S \left[\displaystyle \int dt \left[ U'(t) \right]^{2} \right]^{1/2}}
    = \frac{ P^{1/2} }{ S N_{\phi}^{1/2} \left[\displaystyle \int dt \left[ U'(t) \right]^{2} \right]^{1/2} }.
\end{equation}

\noindent where we have converted to a form with $N_{\phi} = P/\Delta t$ phase bins sampled across the pulse period $P$. As in \cite{Lam_2016}, we define an effective width $W_{\mathrm{eff}}$ such that $\sigma_{\mathrm{S/N}} = {W_{\mathrm{eff}}}/{S N_{\phi}^{1/2}}$, which implies:

\begin{equation}\label{eq:effwidth}
    W_{\mathrm{eff}} = \frac{P^{1/2}}{\displaystyle \left[\int dt \left[ U'(t)\right]^{2} \right]^{1/2}} \rightarrow \displaystyle \left[\int dt \left[ U'(t)\right]^{2} \right]^{1/2} = \frac{P^{1/2}}{W_{\mathrm{eff}}}.
\end{equation} 

We now consider the case where the pulsar emits in both high- and low-energy states (labeled $h$ and $l$, respectively), which we will denote with template shapes $U_{h}(t)$ and $U_{l}(t)$. The template shape of all pulses, $U_{a}(t)$, will be weighted by the respective single pulse intensities of the two states, $I_{h}$ and $I_{l}$, as well as the number of pulses contributing to the total shape, $N_{h}$ and $N_{l}$. Quantitatively, 
\begin{equation}\label{eq:all_pulse}
    U_{a}(t) = \frac{N_{h} I_{h} U_{h}(t) + N_{l} I_{l} U_{l}(t)}{N_{h} I_{h} + N_{l} I_{l}}.
\end{equation}

\noindent The all-pulse average shape can then be calculated by substituting Eq.~\ref{eq:all_pulse} into the left-hand side of Eq.~\ref{eq:effwidth} to obtain
\begin{equation}
    W_{\mathrm{eff}, a} = \frac{P^{1/2}}{\left[ \displaystyle \int dt \left[ \frac{N_{h} I_{h} U'_{h}(t) + N_{l} I_{l} U'_{l}(t)}{N_{h} I_{h} + N_{l} I_{l}} \right]^{2} \right]^{1/2}}.
\end{equation}
\noindent We can expand this expression, writing in $W_{\mathrm{eff},a}^2$ instead of $W_{\mathrm{eff},a}$ for clarity, and replace $\int dt \left[ U'_{i}(t)\right]^{2}=P/W_{\mathrm{eff},i}^2$ using Eq.~\ref{eq:effwidth} to obtain
\begin{eqnarray}\label{eq:approx}
W_{\mathrm{eff},a}^2 & = & \frac{P (N_h I_h + N_l I_l)^2}{\displaystyle{\left[(N_h I_h)^2 \int dt [U_h'(t)]^2 + (N_l I_l)^2 \int dt [U_l'(t)]^2 + 2N_h N_l I_h I_l \int dt U_h'(t) U_l'(t)\right]}} \\
& = & \frac{P\left(\frac{N_h I_h}{N_l I_l}+1\right)^2}{\displaystyle{\left[\left(\frac{N_h I_h}{N_l I_l}\right)^2\frac{P}{W_{\mathrm{eff},h}^2} + \frac{P}{W_{\mathrm{eff},l}^2}+ 2\frac{N_h I_h}{N_l I_l} \int dt U_h'(t) U_l'(t)\right]}}.
\end{eqnarray}

\noindent Therefore, assuming that observationally $N_{h} \ll N_{l}$ (and there are not extreme differences between $I_h$ and $I_l$ so that $N_h I_h / N_l I_l \ll 1$),  we can approximate the above to find
\begin{equation}
\begin{split}
    W_{\mathrm{eff}}^{2} \approx \frac{P}{ {P}/{W_{\mathrm{eff},l}^{2}} } = W_{\mathrm{eff}, l}^{2}.
\end{split}
\end{equation}
\noindent As a result, we obtain that the effective width of the all-pulse template is dominated by the low-fluence-state pulses.

Since the S/N of the individual states are $S_{i}=I_{i} N_{i}^{1/2} / \sigma_{0}$ (by construction/proportionality), assuming that the off-pulse noise at the single pulse level is $\sigma_{0}$ and is the same for both states, then following Eqs.~\ref{eq:sigma_TOA} and \ref{eq:new_sigma_TOA}, the TOA errors for the individual states $i={h,l}$ are
\begin{equation}\label{eq:sigma}
    \sigma_{\mathrm{S/N}, i} = \frac{W_{\mathrm{eff},i}}{S_{i} N_{\phi}^{1/2}} = \frac{W_{\mathrm{eff},i} \sigma_{0}}{I_{i}(N_{i} N_{\phi})^{1/2}}.
\end{equation}
Assuming the peak intensities for both states are roughly aligned in phase, then the single-pulse intensity should be the weighted mean of the individual intensities,
\begin{equation}
    I_a = \frac{N_{h} I_{h} + N_{l} I_{l}}{N_{h} + N_{l}}.
\end{equation}

\noindent Then, we can use Eq.~\ref{eq:sigma} but with the index $i=a$. In the case where $N_{h} \ll N_{l}$, $I_a \approx I_l$, and we can use Eq.~\ref{eq:approx} to replace the effective width $W_{\mathrm{eff},a}$ of the all-pulse template with $W_{\mathrm{eff},l}$ from the low-fluence-state template to find
\begin{equation}
    \sigma_{\mathrm{S/N}, a} = \frac{W_{\mathrm{eff},a} \sigma_{0}}{I_{a}(N_{a} N_{\phi})^{1/2}} \approx \frac{W_{\mathrm{eff},l} \sigma_{0}}{I_{l}(N_{a} N_{\phi})^{1/2}} = \sigma_{\mathrm{S/N},l} \left(\frac{N_l}{N_a}\right)^{1/2}.
\end{equation}

\noindent We see that as expected, using only the low-fluence-state pulses should yield slightly worse timing than using all pulses because, all other factors being equal, the only difference is the number of pulses averaged over given the approximations above. If the effective width of the high-fluence-state template is sufficiently narrow, or the single-pulse intensities sufficiently large, then if we shoud be able to improve upon template-fitting errors (i.e., if $\sigma_{\mathrm{S/N},h} < \sigma_{\mathrm{S/N},a}$ or $\sigma_{\mathrm{S/N},l}$).


\bibliographystyle{mnras}
\bibliography{main}

\begin{thebibliography}{}
\makeatletter
\relax
\def\mn@urlcharsother{\let\do\@makeother \do\$\do\&\do\#\do\^\do\_\do\%\do\~}
\def\mn@doi{\begingroup\mn@urlcharsother \@ifnextchar [ {\mn@doi@}
  {\mn@doi@[]}}
\def\mn@doi@[#1]#2{\def\@tempa{#1}\ifx\@tempa\@empty \href
  {http://dx.doi.org/#2} {doi:#2}\else \href {http://dx.doi.org/#2} {#1}\fi
  \endgroup}
\def\mn@eprint#1#2{\mn@eprint@#1:#2::\@nil}
\def\mn@eprint@arXiv#1{\href {http://arxiv.org/abs/#1} {{\tt arXiv:#1}}}
\def\mn@eprint@dblp#1{\href {http://dblp.uni-trier.de/rec/bibtex/#1.xml}
  {dblp:#1}}
\def\mn@eprint@#1:#2:#3:#4\@nil{\def\@tempa {#1}\def\@tempb {#2}\def\@tempc
  {#3}\ifx \@tempc \@empty \let \@tempc \@tempb \let \@tempb \@tempa \fi \ifx
  \@tempb \@empty \def\@tempb {arXiv}\fi \@ifundefined
  {mn@eprint@\@tempb}{\@tempb:\@tempc}{\expandafter \expandafter \csname
  mn@eprint@\@tempb\endcsname \expandafter{\@tempc}}}

\bibitem[\protect\citeauthoryear{{Agazie} et~al.,}{{Agazie}
  et~al.}{2023}]{2023ApJ...951L...8A}
{Agazie} G.,  et~al., 2023, \mn@doi [\apjl] {10.3847/2041-8213/acdac6}, \href
  {https://ui.adsabs.harvard.edu/abs/2023ApJ...951L...8A} {951, L8}

\bibitem[\protect\citeauthoryear{Aggarwal et~al.,}{Aggarwal
  et~al.}{2020}]{Aggarwal2020}
Aggarwal K.,  et~al., 2020, \mn@doi [Journal of Open Source Software]
  {10.21105/joss.02750}, 5, 2750

\bibitem[\protect\citeauthoryear{{Alam} et~al.,}{{Alam}
  et~al.}{2021}]{2021ApJS..252....4A}
{Alam} M.~F.,  et~al., 2021, \mn@doi [\apjs] {10.3847/1538-4365/abc6a0}, \href
  {https://ui.adsabs.harvard.edu/abs/2021ApJS..252....4A} {252, 4}

\bibitem[\protect\citeauthoryear{Ankerst, Breunig, Kriegel  \& Sander}{Ankerst
  et~al.}{1999}]{ankerst1999optics}
Ankerst M.,  Breunig M.~M.,  Kriegel H.-P.,   Sander J.,  1999, ACM Sigmod
  record, 28, 49

\bibitem[\protect\citeauthoryear{{Antoniadis} et~al.,}{{Antoniadis}
  et~al.}{2013}]{2013Sci...340..448A}
{Antoniadis} J.,  et~al., 2013, \mn@doi [Science] {10.1126/science.1233232},
  \href {https://ui.adsabs.harvard.edu/abs/2013Sci...340..448A} {340, 448}

\bibitem[\protect\citeauthoryear{{Arzoumanian} et~al.,}{{Arzoumanian}
  et~al.}{2016}]{2016ApJ...821...13A}
{Arzoumanian} Z.,  et~al., 2016, \mn@doi [\apj] {10.3847/0004-637X/821/1/13},
  \href {https://ui.adsabs.harvard.edu/abs/2016ApJ...821...13A} {821, 13}

\bibitem[\protect\citeauthoryear{{Arzoumanian} et~al.,}{{Arzoumanian}
  et~al.}{2020}]{2020ApJ...905L..34A}
{Arzoumanian} Z.,  et~al., 2020, \mn@doi [\apjl] {10.3847/2041-8213/abd401},
  \href {https://ui.adsabs.harvard.edu/abs/2020ApJ...905L..34A} {905, L34}

\bibitem[\protect\citeauthoryear{{Astropy Collaboration} et~al.,}{{Astropy
  Collaboration} et~al.}{2013}]{astropy:2013}
{Astropy Collaboration} et~al., 2013, \mn@doi [\aap]
  {10.1051/0004-6361/201322068}, \href
  {https://ui.adsabs.harvard.edu/abs/2013A&A...558A..33A} {558, A33}

\bibitem[\protect\citeauthoryear{{Astropy Collaboration} et~al.,}{{Astropy
  Collaboration} et~al.}{2018}]{astropy:2018}
{Astropy Collaboration} et~al., 2018, \mn@doi [\aj] {10.3847/1538-3881/aabc4f},
  \href {https://ui.adsabs.harvard.edu/abs/2018AJ....156..123A} {156, 123}

\bibitem[\protect\citeauthoryear{{Astropy Collaboration} et~al.,}{{Astropy
  Collaboration} et~al.}{2022}]{astropy:2022}
{Astropy Collaboration} et~al., 2022, \mn@doi [\apj]
  {10.3847/1538-4357/ac7c74}, \href
  {https://ui.adsabs.harvard.edu/abs/2022ApJ...935..167A} {935, 167}

\bibitem[\protect\citeauthoryear{{Backer}}{{Backer}}{1975}]{1973ApJ...182..245B}
{Backer} D.~C.,  1975, \mn@doi [\apj] {10.1086/152134}, \href
  {https://ui.adsabs.harvard.edu/abs/1973ApJ...182..245B} {182, 245}

\bibitem[\protect\citeauthoryear{{Bailes} et~al.,}{{Bailes}
  et~al.}{1994}]{1994ApJ...425L..41B}
{Bailes} M.,  et~al., 1994, \mn@doi [\apjl] {10.1086/187306}, \href
  {https://ui.adsabs.harvard.edu/abs/1994ApJ...425L..41B} {425, L41}

\bibitem[\protect\citeauthoryear{Bellman}{Bellman}{2003}]{bellman2003dynamic}
Bellman R.,  2003, Dynamic Programming.
Dover Books on Computer Science Series, Dover Publications, \url
  {https://books.google.com/books?id=fyVtp3EMxasC}

\bibitem[\protect\citeauthoryear{{Cheng} \& {Ruderman}}{{Cheng} \&
  {Ruderman}}{1980}]{1980ApJ...235..576C}
{Cheng} A.~F.,  {Ruderman} M.~A.,  1980, \mn@doi [\apj] {10.1086/157661}, \href
  {https://ui.adsabs.harvard.edu/abs/1980ApJ...235..576C} {235, 576}

\bibitem[\protect\citeauthoryear{Comaniciu \& Meer}{Comaniciu \&
  Meer}{2002}]{1000236}
Comaniciu D.,  Meer P.,  2002, \mn@doi [IEEE Transactions on Pattern Analysis
  and Machine Intelligence] {10.1109/34.1000236}, 24, 603

\bibitem[\protect\citeauthoryear{{Cordes} \& {Downs}}{{Cordes} \&
  {Downs}}{1985}]{1985ApJS...59..343C}
{Cordes} J.~M.,  {Downs} G.~S.,  1985, \mn@doi [\apjs] {10.1086/191076}, \href
  {https://ui.adsabs.harvard.edu/abs/1985ApJS...59..343C} {59, 343}

\bibitem[\protect\citeauthoryear{{Cordes} \& {Shannon}}{{Cordes} \&
  {Shannon}}{2010}]{2010arXiv1010.3785C}
{Cordes} J.~M.,  {Shannon} R.~M.,  2010, \mn@doi [arXiv e-prints]
  {10.48550/arXiv.1010.3785}, \href
  {https://ui.adsabs.harvard.edu/abs/2010arXiv1010.3785C} {p. arXiv:1010.3785}

\bibitem[\protect\citeauthoryear{{Cordes}, {Wolszczan}, {Dewey}, {Blaskiewicz}
  \& {Stinebring}}{{Cordes} et~al.}{1990}]{1990ApJ...349..245C}
{Cordes} J.~M.,  {Wolszczan} A.,  {Dewey} R.~J.,  {Blaskiewicz} M.,
  {Stinebring} D.~R.,  1990, \mn@doi [\apj] {10.1086/168310}, \href
  {https://ui.adsabs.harvard.edu/abs/1990ApJ...349..245C} {349, 245}

\bibitem[\protect\citeauthoryear{Craft}{Craft}{1970}]{craft_1970}
Craft H.~D.,  1970, PhD thesis, Cornell University

\bibitem[\protect\citeauthoryear{{Cromartie} et~al.,}{{Cromartie}
  et~al.}{2020}]{2020NatAs...4...72C}
{Cromartie} H.~T.,  et~al., 2020, \mn@doi [Nature Astronomy]
  {10.1038/s41550-019-0880-2}, \href
  {https://ui.adsabs.harvard.edu/abs/2020NatAs...4...72C} {4, 72}

\bibitem[\protect\citeauthoryear{Demorest et~al.,}{Demorest
  et~al.}{2012}]{Demorest_2013}
Demorest P.~B.,  et~al., 2012, \mn@doi [The Astrophysical Journal]
  {10.1088/0004-637X/762/2/94}, 762, 94

\bibitem[\protect\citeauthoryear{{Demorest} et~al.,}{{Demorest}
  et~al.}{2013}]{2013ApJ...762...94D}
{Demorest} P.~B.,  et~al., 2013, \mn@doi [\apj] {10.1088/0004-637X/762/2/94},
  \href {https://ui.adsabs.harvard.edu/abs/2013ApJ...762...94D} {762, 94}

\bibitem[\protect\citeauthoryear{{Detweiler}}{{Detweiler}}{1979}]{1979ApJ...234.1100D}
{Detweiler} S.,  1979, \mn@doi [\apj] {10.1086/157593}, \href
  {https://ui.adsabs.harvard.edu/abs/1979ApJ...234.1100D} {234, 1100}

\bibitem[\protect\citeauthoryear{Dolch et~al.,}{Dolch
  et~al.}{2014}]{Dolch_2014}
Dolch T.,  et~al., 2014, \mn@doi [The Astrophysical Journal]
  {10.1088/0004-637X/794/1/21}, 794, 21

\bibitem[\protect\citeauthoryear{{Downs} \& {Reichley}}{{Downs} \&
  {Reichley}}{1983}]{1983ApJS...53..169D}
{Downs} G.~S.,  {Reichley} P.~E.,  1983, \mn@doi [\apjs] {10.1086/190890},
  \href {https://ui.adsabs.harvard.edu/abs/1983ApJS...53..169D} {53, 169}

\bibitem[\protect\citeauthoryear{{Edwards} \& {Stappers}}{{Edwards} \&
  {Stappers}}{2003}]{2003A&A...407..273E}
{Edwards} R.~T.,  {Stappers} B.~W.,  2003, \mn@doi [\aap]
  {10.1051/0004-6361:20030716}, \href
  {https://ui.adsabs.harvard.edu/abs/2003A&A...407..273E} {407, 273}

\bibitem[\protect\citeauthoryear{Ester, Kriegel, Sander  \& Xu}{Ester
  et~al.}{1996}]{Ester1996ADA}
Ester M.,  Kriegel H.-P.,  Sander J.,   Xu X.,  1996, in Knowledge Discovery
  and Data Mining. \url {https://api.semanticscholar.org/CorpusID:355163}

\bibitem[\protect\citeauthoryear{{Fonseca} et~al.,}{{Fonseca}
  et~al.}{2016}]{2016ApJ...832..167F}
{Fonseca} E.,  et~al., 2016, \mn@doi [\apj] {10.3847/0004-637X/832/2/167},
  \href {https://ui.adsabs.harvard.edu/abs/2016ApJ...832..167F} {832, 167}

\bibitem[\protect\citeauthoryear{{Fonseca} et~al.,}{{Fonseca}
  et~al.}{2021}]{2021ApJ...915L..12F}
{Fonseca} E.,  et~al., 2021, \mn@doi [\apjl] {10.3847/2041-8213/ac03b8}, \href
  {https://ui.adsabs.harvard.edu/abs/2021ApJ...915L..12F} {915, L12}

\bibitem[\protect\citeauthoryear{{Gancio} et~al.,}{{Gancio}
  et~al.}{2020}]{2020A&A...633A..84G}
{Gancio} G.,  et~al., 2020, \mn@doi [\aap] {10.1051/0004-6361/201936525}, \href
  {https://ui.adsabs.harvard.edu/abs/2020A&A...633A..84G} {633, A84}

\bibitem[\protect\citeauthoryear{Hartung, Knapp  \& Sinha}{Hartung
  et~al.}{2011}]{hartung2011statistical}
Hartung J.,  Knapp G.,   Sinha B.,  2011, Statistical Meta-Analysis with
  Applications.
Wiley Series in Probability and Statistics, Wiley, \url
  {https://books.google.com/books?id=JEoNB_2NONQC}

\bibitem[\protect\citeauthoryear{{Hassall} et~al.,}{{Hassall}
  et~al.}{2012}]{2012A&A...543A..66H}
{Hassall} T.~E.,  et~al., 2012, \mn@doi [\aap] {10.1051/0004-6361/201218970},
  \href {https://ui.adsabs.harvard.edu/abs/2012A&A...543A..66H} {543, A66}

\bibitem[\protect\citeauthoryear{{Hellings} \& {Downs}}{{Hellings} \&
  {Downs}}{1983}]{1983ApJ...265L..39H}
{Hellings} R.~W.,  {Downs} G.~S.,  1983, \mn@doi [\apjl] {10.1086/183954},
  \href {https://ui.adsabs.harvard.edu/abs/1983ApJ...265L..39H} {265, L39}

\bibitem[\protect\citeauthoryear{Hemberger \& Stinebring}{Hemberger \&
  Stinebring}{2008}]{Hemberger_2008}
Hemberger D.~A.,  Stinebring D.~R.,  2008, \mn@doi [The Astrophysical Journal]
  {10.1086/528985}, 674, L37

\bibitem[\protect\citeauthoryear{{Hotan}, {van Straten}  \&
  {Manchester}}{{Hotan} et~al.}{2004}]{2004PASA...21..302H}
{Hotan} A.~W.,  {van Straten} W.,   {Manchester} R.~N.,  2004, \mn@doi [\pasa]
  {10.1071/AS04022}, \href
  {https://ui.adsabs.harvard.edu/abs/2004PASA...21..302H} {21, 302}

\bibitem[\protect\citeauthoryear{{Kerr}}{{Kerr}}{2015}]{2015MNRAS.452..607K}
{Kerr} M.,  2015, \mn@doi [\mnras] {10.1093/mnras/stv1296}, \href
  {https://ui.adsabs.harvard.edu/abs/2015MNRAS.452..607K} {452, 607}

\bibitem[\protect\citeauthoryear{{Kloumann} \& {Rankin}}{{Kloumann} \&
  {Rankin}}{2011}]{2011AAS...21713907K}
{Kloumann} I.~M.,  {Rankin} J.~M.,  2011, in American Astronomical Society
  Meeting Abstracts \#217. p. 139.07

\bibitem[\protect\citeauthoryear{{Kramer} et~al.,}{{Kramer}
  et~al.}{2006}]{2006Sci...314...97K}
{Kramer} M.,  et~al., 2006, \mn@doi [Science] {10.1126/science.1132305}, \href
  {https://ui.adsabs.harvard.edu/abs/2006Sci...314...97K} {314, 97}

\bibitem[\protect\citeauthoryear{{Krishnamohan} \& {Downs}}{{Krishnamohan} \&
  {Downs}}{1983}]{1983ApJ...265..372K}
{Krishnamohan} S.,  {Downs} G.~S.,  1983, \mn@doi [\apj] {10.1086/160682},
  \href {https://ui.adsabs.harvard.edu/abs/1983ApJ...265..372K} {265, 372}

\bibitem[\protect\citeauthoryear{{Lam}}{{Lam}}{2017}]{2017ascl.soft06011L}
{Lam} M.~T.,  2017, {PyPulse: PSRFITS handler}, Astrophysics Source Code
  Library, record ascl:1706.011 (\mn@eprint {ascl} {1706.011})

\bibitem[\protect\citeauthoryear{Lam et~al.,}{Lam et~al.}{2016}]{Lam_2016}
Lam M.~T.,  et~al., 2016, \mn@doi [The Astrophysical Journal]
  {10.3847/0004-637X/819/2/155}, 819, 155

\bibitem[\protect\citeauthoryear{Lam et~al.,}{Lam et~al.}{2019}]{Lam_2019}
Lam M.~T.,  et~al., 2019, \mn@doi [The Astrophysical Journal]
  {10.3847/1538-4357/ab01cd}, 872, 193

\bibitem[\protect\citeauthoryear{{Lazarus}, {Karuppusamy}, {Graikou},
  {Caballero}, {Champion}, {Lee}, {Verbiest}  \& {Kramer}}{{Lazarus}
  et~al.}{2020}]{2020ascl.soft03008L}
{Lazarus} P.,  {Karuppusamy} R.,  {Graikou} E.,  {Caballero} R.~N.,  {Champion}
  D.~J.,  {Lee} K.~J.,  {Verbiest} J.~P.~W.,   {Kramer} M.,  2020, {CoastGuard:
  Automated timing data reduction pipeline}, Astrophysics Source Code Library,
  record ascl:2003.008

\bibitem[\protect\citeauthoryear{Liu, Keane, Lee, Kramer, Cordes  \&
  Purver}{Liu et~al.}{2012}]{10.1111/j.1365-2966.2011.20041.x}
Liu K.,  Keane E.~F.,  Lee K.~J.,  Kramer M.,  Cordes J.~M.,   Purver M.~B.,
  2012, \mn@doi [Monthly Notices of the Royal Astronomical Society]
  {10.1111/j.1365-2966.2011.20041.x}, 420, 361

\bibitem[\protect\citeauthoryear{{L{\"o}hmer}, {Kramer}, {Driebe}, {Jessner},
  {Mitra}  \& {Lyne}}{{L{\"o}hmer} et~al.}{2004}]{2004A&A...426..631L}
{L{\"o}hmer} O.,  {Kramer} M.,  {Driebe} T.,  {Jessner} A.,  {Mitra} D.,
  {Lyne} A.~G.,  2004, \mn@doi [\aap] {10.1051/0004-6361:20041031}, \href
  {https://ui.adsabs.harvard.edu/abs/2004A&A...426..631L} {426, 631}

\bibitem[\protect\citeauthoryear{Lommen \& Demorest}{Lommen \&
  Demorest}{2013}]{Lommen_2013}
Lommen A.~N.,  Demorest P.,  2013, \mn@doi [Classical and Quantum Gravity]
  {10.1088/0264-9381/30/22/224001}, 30, 224001

\bibitem[\protect\citeauthoryear{{Lousto} et~al.,}{{Lousto}
  et~al.}{2022}]{2022MNRAS.509.5790L}
{Lousto} C.~O.,  et~al., 2022, \mn@doi [\mnras] {10.1093/mnras/stab3287}, \href
  {https://ui.adsabs.harvard.edu/abs/2022MNRAS.509.5790L} {509, 5790}

\bibitem[\protect\citeauthoryear{{Luo} et~al.,}{{Luo} et~al.}{2021}]{PINT}
{Luo} J.,  et~al., 2021, \mn@doi [\apj] {10.3847/1538-4357/abe62f}, \href
  {https://ui.adsabs.harvard.edu/abs/2021ApJ...911...45L} {911, 45}

\bibitem[\protect\citeauthoryear{{Maan}, {van Leeuwen}  \& {Vohl}}{{Maan}
  et~al.}{2021}]{2021A&A...650A..80M}
{Maan} Y.,  {van Leeuwen} J.,   {Vohl} D.,  2021, \mn@doi [\aap]
  {10.1051/0004-6361/202040164}, \href
  {https://ui.adsabs.harvard.edu/abs/2021A&A...650A..80M} {650, A80}

\bibitem[\protect\citeauthoryear{MacQueen et~al.}{MacQueen
  et~al.}{1967}]{macqueen1967some}
MacQueen J.,  et~al., 1967, in Proceedings of the fifth Berkeley symposium on
  mathematical statistics and probability. pp 281--297

\bibitem[\protect\citeauthoryear{{McKinnon} \& {Hankins}}{{McKinnon} \&
  {Hankins}}{1993}]{1993A&A...269..325M}
{McKinnon} M.~M.,  {Hankins} T.~H.,  1993, \aap, \href
  {https://ui.adsabs.harvard.edu/abs/1993A&A...269..325M} {269, 325}

\bibitem[\protect\citeauthoryear{{Morello} et~al.,}{{Morello}
  et~al.}{2019}]{2019MNRAS.483.3673M}
{Morello} V.,  et~al., 2019, \mn@doi [\mnras] {10.1093/mnras/sty3328}, \href
  {https://ui.adsabs.harvard.edu/abs/2019MNRAS.483.3673M} {483, 3673}

\bibitem[\protect\citeauthoryear{Newville, Stensitzki, Allen  \&
  Ingargiola}{Newville et~al.}{2014}]{LMFIT}
Newville M.,  Stensitzki T.,  Allen D.~B.,   Ingargiola A.,  2014, {LMFIT:
  Non-Linear Least-Square Minimization and Curve-Fitting for Python},
  \mn@doi{10.5281/zenodo.11813}, \url {https://doi.org/10.5281/zenodo.11813}

\bibitem[\protect\citeauthoryear{{Os\l{}owski}, {van Straten}, {Bailes},
  {Jameson}  \& {Hobbs}}{{Os\l{}owski} et~al.}{2014}]{2014MNRAS.441.3148O}
{Os\l{}owski} S.,  {van Straten} W.,  {Bailes} M.,  {Jameson} A.,   {Hobbs} G.,
   2014, \mn@doi [\mnras] {10.1093/mnras/stu804}, \href
  {https://ui.adsabs.harvard.edu/abs/2014MNRAS.441.3148O} {441, 3148}

\bibitem[\protect\citeauthoryear{Pedregosa et~al.,}{Pedregosa
  et~al.}{2011}]{scikit-learn}
Pedregosa F.,  et~al., 2011, Journal of Machine Learning Research, 12, 2825

\bibitem[\protect\citeauthoryear{{Phillips} \& {Wolszczan}}{{Phillips} \&
  {Wolszczan}}{1992}]{1992ApJ...385..273P}
{Phillips} J.~A.,  {Wolszczan} A.,  1992, \mn@doi [\apj] {10.1086/170935},
  \href {https://ui.adsabs.harvard.edu/abs/1992ApJ...385..273P} {385, 273}

\bibitem[\protect\citeauthoryear{{Pilia} et~al.,}{{Pilia}
  et~al.}{2016}]{2016yCat..35860092P}
{Pilia} M.,  et~al., 2016, VizieR Online Data Catalog, \href
  {https://ui.adsabs.harvard.edu/abs/2016yCat..35860092P} {pp J/A+A/586/A92}

\bibitem[\protect\citeauthoryear{{Sander}, {Ester}, {Kriegel}  \&
  {Xu}}{{Sander} et~al.}{1998}]{1998DMKD....2..169S}
{Sander} J.,  {Ester} M.,  {Kriegel} H.-P.,   {Xu} X.,  1998, \mn@doi [Data
  Mining and Knowledge Discovery] {10.1023/A:1009745219419}, \href
  {https://ui.adsabs.harvard.edu/abs/1998DMKD....2..169S} {2, 169}

\bibitem[\protect\citeauthoryear{Schubert \& Gertz}{Schubert \&
  Gertz}{2018}]{Schubert2018ImprovingTC}
Schubert E.,  Gertz M.,  2018, in Lernen, Wissen, Daten, Analysen. \url
  {https://api.semanticscholar.org/CorpusID:52088333}

\bibitem[\protect\citeauthoryear{Schubert, Sander, Ester, Kriegel  \&
  Xu}{Schubert et~al.}{2017}]{10.1145/3068335}
Schubert E.,  Sander J.,  Ester M.,  Kriegel H.~P.,   Xu X.,  2017, \mn@doi
  [ACM Trans. Database Syst.] {10.1145/3068335}, 42

\bibitem[\protect\citeauthoryear{Shahar}{Shahar}{2017}]{weighted_variance}
Shahar D.,  2017, \mn@doi [Open Journal of Statistics]
  {10.4236/ojs.2017.72017}, 7, 216

\bibitem[\protect\citeauthoryear{Shannon \& Cordes}{Shannon \&
  Cordes}{2010}]{Shannon_2010}
Shannon R.~M.,  Cordes J.~M.,  2010, \mn@doi [The Astrophysical Journal]
  {10.1088/0004-637X/725/2/1607}, 725, 1607

\bibitem[\protect\citeauthoryear{{Shannon} \& {Cordes}}{{Shannon} \&
  {Cordes}}{2012}]{2012ApJ...761...64S}
{Shannon} R.~M.,  {Cordes} J.~M.,  2012, \mn@doi [\apj]
  {10.1088/0004-637X/761/1/64}, \href
  {https://ui.adsabs.harvard.edu/abs/2012ApJ...761...64S} {761, 64}

\bibitem[\protect\citeauthoryear{Shannon et~al.,}{Shannon
  et~al.}{2014a}]{10.1093/mnras/stu1213}
Shannon R.~M.,  et~al., 2014a, \mn@doi [Monthly Notices of the Royal
  Astronomical Society] {10.1093/mnras/stu1213}, 443, 1463

\bibitem[\protect\citeauthoryear{{Shannon} et~al.,}{{Shannon}
  et~al.}{2014b}]{2014MNRAS.443.1463S}
{Shannon} R.~M.,  et~al., 2014b, \mn@doi [\mnras] {10.1093/mnras/stu1213},
  \href {https://ui.adsabs.harvard.edu/abs/2014MNRAS.443.1463S} {443, 1463}

\bibitem[\protect\citeauthoryear{{Taylor}}{{Taylor}}{1992}]{1992RSPTA.341..117T}
{Taylor} J.~H.,  1992, \mn@doi [Philosophical Transactions of the Royal Society
  of London Series A] {10.1098/rsta.1992.0088}, \href
  {https://ui.adsabs.harvard.edu/abs/1992RSPTA.341..117T} {341, 117}

\bibitem[\protect\citeauthoryear{{Teixeira}, {Kloumann}  \& M.}{{Teixeira}
  et~al.}{2012}]{TeixeiraProfileSI}
{Teixeira} M.~M.,  {Kloumann} I.~M.,   M. R.~J.,  2012. \url
  {https://api.semanticscholar.org/CorpusID:11012461}

\bibitem[\protect\citeauthoryear{{Wolszczan} \& {Frail}}{{Wolszczan} \&
  {Frail}}{1992}]{1992Natur.355..145W}
{Wolszczan} A.,  {Frail} D.~A.,  1992, \mn@doi [\nat] {10.1038/355145a0}, \href
  {https://ui.adsabs.harvard.edu/abs/1992Natur.355..145W} {355, 145}

\bibitem[\protect\citeauthoryear{{Zubieta} et~al.,}{{Zubieta}
  et~al.}{2023}]{2023MNRAS.521.4504Z}
{Zubieta} E.,  et~al., 2023, \mn@doi [\mnras] {10.1093/mnras/stad723}, \href
  {https://ui.adsabs.harvard.edu/abs/2023MNRAS.521.4504Z} {521, 4504}

\bibitem[\protect\citeauthoryear{{van Straten} \& {Bailes}}{{van Straten} \&
  {Bailes}}{2011}]{2011PASA...28....1V}
{van Straten} W.,  {Bailes} M.,  2011, \mn@doi [\pasa] {10.1071/AS10021}, \href
  {https://ui.adsabs.harvard.edu/abs/2011PASA...28....1V} {28, 1}

\makeatother
\end{thebibliography}



\end{document}